\documentclass{sig-alternate}
\usepackage{multirow, rotating, amsmath, wasysym, paralist}
\usepackage[tight]{subfigure}
\usepackage[ruled,linesnumbered,vlined]{algorithm2e}
\mathchardef\Gamma="0100 \mathchardef\Delta="0101
\mathchardef\Theta="0102 \mathchardef\Lambda="0103
\mathchardef\Xi="0104 \mathchardef\Pi="0105
\mathchardef\Sigma="0106 \mathchardef\Upsilon="0107
\mathchardef\Phi="0108 \mathchardef\Psi="0109
\mathchardef\Omega="010A

\newcommand{\outline}[1]{}

\usepackage{cite}
\usepackage{xspace}
\usepackage{url}
\usepackage{graphicx}
\usepackage{latexsym}
\usepackage{amssymb}
\usepackage{amsfonts}
\usepackage{psfrag}
\usepackage{wrapfig}
\usepackage{comment}
\usepackage{alltt}
\usepackage{color}

\newtheorem{defn}{Definition}[section]

\newtheorem{thm}{Theorem}

\newcommand{\ie}{\emph{i.e.}\xspace}
\newcommand{\eg}{\emph{e.g.}\xspace}

\newcommand{\etal}{\frenchspacing{}\emph{et al{.}}\xspace}

\newtheorem{lemma}{Lemma}[section]
\newtheorem{corollary}{Corollary}[section]

\newcommand{\Comment}[1]{}

\def \R {\mathbb{R}}
\def \N {\mathcal{N}}
\def \Ah {\widehat{A}}

\def \Ph {\widehat{P}}
\def \Pt {\widetilde{P}}
\def \u {\mathbf{u}}
\def \ut {\widetilde{\u}}
\def \uh {\widehat{\u}}
\def \Er {\mathcal{E}}
\def \Uh {\widehat{U}}
\def \Ut {\widetilde{U}}

\title{A Random Matrix Approach to Differential Privacy and Structure Preserved Social Network Graph Publishing}

\numberofauthors{1}
\author{
\alignauthor
Faraz Ahmed, Rong Jin and Alex X. Liu\\
       \affaddr{Department of Computer Science and Engineering}\\
       \affaddr{Michigan State University}\\
       \affaddr{East Lansing, Michigan, USA}\\
       \email{\{farazah, rongjin, alexliu\}@cse.msu.edu}
}
\begin{document}

\maketitle
\begin{abstract}
Online social networks are being increasingly used for analyzing
various societal phenomena such as epidemiology, information
dissemination, marketing and sentiment flow.
Popular analysis techniques such as clustering and influential node
analysis, require the computation of eigenvectors of the real
graph's adjacency matrix.
Recent de-anonymization attacks on Netflix and AOL datasets show
that an open access to such graphs pose privacy threats.
Among the various privacy preserving models, \textit{Differential
privacy} provides the strongest privacy guarantees.

In this paper we propose a privacy preserving mechanism for
publishing social network graph data, which satisfies differential
privacy guarantees by utilizing a combination of theory of random
matrix and that of differential privacy.
The key idea is to project each row of an adjacency matrix to a low
dimensional space using the random projection approach and then
perturb the projected matrix with random noise.
We show that as compared to existing approaches for differential
private approximation of eigenvectors, our approach is
computationally efficient, preserves the utility and satisfies
differential privacy.
We evaluate our approach on social network graphs of Facebook, Live
Journal and Pokec.
The results show that even for high values of noise variance
$\sigma=1$ the clustering quality given by normalized mutual
information gain is as low as $0.74$.
For influential node discovery, the propose approach is able to
correctly recover $80\%$ of the most influential nodes.
We also compare our results with an approach presented in
\cite{wangdifferential}, which directly perturbs the eigenvector of
the original data by a Laplacian noise.
The results show that this approach requires a large random
perturbation in order to preserve the differential privacy, which
leads to a poor estimation of eigenvectors for large social
networks.
\end{abstract}

\vspace{-0.07in}

{\sloppy
\section{Introduction} \label{sec: introduction}

\subsection{Background and Motivation} \label{subsec:background}
Online Social Networks (OSNs) have become an essential part of
modern life.
Billions of users connect and share information using OSNs such as
Facebook and Twitter.
Graphs obtained from these OSNs can provide useful insights on
various fundamental societal phenomena such as epidemiology,
information dissemination, marketing, and sentiment flow
\cite{ahn2007analysis, rogers1995diffusion, gruhl2004information,
domingos2001mining, richardson2002mining}.
Various analysis methods \cite{girvan2002community, du2007community,
cha2009measurement, kempe2003maximizing, kwak2010twitter} have been
applied to OSNs by explicitly exploring its graph structure, such as
clustering analysis for automatically identifying online communities
and node influence analysis for recognizing the influential nodes in
social networks.
The basis of all these analysis is to represent a social network
graph by an adjacency matrix and then represent individual nodes by
vectors derived from the top eigenvectors of the adjacency matrix.
Thus, all these analysis methods require real social network graphs.

Unfortunately, OSNs often refuse to publish their social network
graphs due to privacy concerns.
Social network graphs contain sensitive information about
individuals such as an user's topological characteristics in a graph
(\eg, number of social ties, influence in a community, etc).
From the user perspective, the sensitive information revealed from a
social network graph can be exploited in many ways such as the
propagation of malware and spam \cite{thomas2010koobface}.
From the OSN perspective, disclosing sensitive user information put
them in the risk of violating privacy laws.
A natural way to bridge the gap is to anonymize original social
network graphs (by means such as removing identifiers) and publish
the anonymized ones.
For example, Netflix published anonymized movie ratings of 500,000
subscribers and AOL published search queries of 658,000 users
\cite{hansell2006aol, narayanan2008robust}.
However, such anonymization is vulnerable to privacy attacks
\cite{backstrom2007wherefore, narayanan2009anonymizing} where
attackers can identify personal information by linking two or more
separately innocuous databases.
For example, recently, de-anonymization attacks were successful on
Netflix and AOL datasets, which resulted in Netflix and AOL being
sued \cite{hansell2006aol, narayanan2008robust}.

\subsection{Problem Statement}\label{subsec:problem}
In this paper, we aim to develop a scheme for publishing social
network graphs with differential privacy guarantees.
The concept of differential privacy was raised in the context of
statistical database, where a trusted party holds a dataset $D$
containing sensitive information (\eg medical records) and wants to
publish a dataset $D'$ that provides the same global statistical
information as $D$ while preserving the privacy information of each
individual user.
%
%
Recently, differential privacy has become the widely accepted
criteria for privacy preserving data publishing because it provides
robust privacy guarantees for publishing sensitive data
\cite{dwork2006differential, dwork2008differential,
dwork2009differential}.

This privacy preserving social graph publishing scheme should
satisfy the following two requirements.
First, the published data should maintain the utility of the
original data.
As many analysis of social networks are based on the top
eigenvectors of the adjacency matrices derived from social networks,
the utility of the published data will be measured by how well the
top eigenvectors of the published data can be approximated to the
eigenvectors of the original data.
Second, the scheme should achieve the desired privacy guarantees,
\ie, an adversary should learn nothing more about any individual
from the published data, regardless of the presence or absence of an
individual's record in the data.
We emphasize that these two goals are often conflicting: to preserve
the differential privacy of individuals, a sufficiently large amount
of random noise has to be added to the published data, which could
potentially result in a large error in approximating the top
eigenvectors of the original data.
Our goal is to achieve a best tradeoff between privacy and utility.

\subsection{Limitations of Prior Art}\label{subsec:limitations}


A few schemes have been developed to approximate eigenvectors and
eigenvalues of matrices in a differential private manner
\cite{hardt2012beyond}
\cite{chaudhuri2012near,kapralov2012differentially}.
Their main idea is to perturb the original matrices by adding random
noise and then publish the perturbed matrices.
The key limitation of this approach is that given $n$ users in the
social network, they have to publish a large \emph{dense} matrix of
size $n\times n$, leading to a high cost in both computation and
storage space.
Recently, Wang \etal proposed to perturb the eigenvectors of the
original matrices by adding random noises and then publish the
perturbed eigenvectors \cite{wangdifferential}.
If we are interested in the first $k$ eigenvectors of the adjacency
matrix, where $k \ll n$, we only need to publish a matrix of size
$n\times k$.
Although this reduces computation cost and storage space, it
requires a large amount of random perturbation in order to preserve
differential privacy, which leads to poor estimation of eigenvectors
for large social networks.

\subsection{Proposed Approach}\label{subsec:methodology}
We propose a random matrix approach to address the above limitations
by leveraging the theories of random matrix and differential
privacy.
Our key idea is to first project each row of an adjacency matrix
into a low dimensional space using random projection, and then
perturb the projected matrix with random noise, and finally publish
the perturbed and projected matrix.
The random projection is critical in our approach.
First, it reduces the dimensionality of the matrix to be published,
avoiding the difficulty of publishing a large dense matrix.
Second, according to the theory of random matrix
\cite{Halko:2011:FSR}, the random projection step allows us to
preserve the top eigenvectors of the adjacency matrix.
Third, the random projection step by itself has the ability of
achieving differential privacy, which makes it possible to ensure
differential privacy in the second step by introducing a
\emph{small} random perturbation \cite{proserpio2012workflow,
blocki2012johnson}.

\subsection{Validation of Proposed Approach}\label{subsec:methodology}
To validate our differential private random matrix approach and to
illustrate the utility preservation of eigen-spectrum, we perform
experiments over graphs obtained from Facebook, Live Journal and
Pokec social networks.
We analyze the impact of perturbation by evaluating the utility of
the published data for two different applications which require
spectral information of a graph.
First, we consider clustering of social networks, which has been
widely used for community detection in social networks. We choose
spectral clustering algorithm in our study, which depends on the
eigenvectors of the adjacency matrix. Next, we examine the problem
of identifying the ranks of influential nodes in a social network
graph.

\subsection{Key Contributions} \label{sec:contributions}
We make three key contributions in this paper.
First, we propose a random projection approach which utilizes random
matrix theory to reduce the dimensions of the adjacency matrix and
achieves differential privacy by adding small amount of noise.
As online social networks consists of millions or even billions of
nodes, it is crucial to minimize computational cost and storage
space.
The dimensionality reduction reduces the computational cost of the
algorithm and small noise addition maintains the utility of the
data.
Second, we formally prove that our scheme achieves differential
privacy.
We also provide theoretical error bounds for approximating top$-k$
eigenvectors.
Finally, we perform evaluation by analyzing the utility of the
published data for two different applications which require spectral
information of a graph. We consider clustering of social networks
and the problem of identifying the ranks of influential nodes in a
social network graph. We also compare our results with an approach
presented in \cite{wangdifferential}, which directly perturbs the
eigenvector of the original data by a Laplacian noise.

\section{Related Work} \label{sec:related}
\subsection{Differential Privacy}\label{sec:differential}
The seminal work of D. Work et. al \cite{dwork2006differential}, on
differential privacy provides formal privacy guarantees that do not
depend on an adversary's background knowledge.
The notion of differential privacy was developed through a series of
research work presented in \cite{dwork2006calibrating,
blum2005practical, kenthapadi2012privacy}.
Popular differential private mechanisms which are used in publishing
sensitive data include Laplace mechanism \cite{dwork2006calibrating}
and the Exponential mechanism \cite{mcsherry2007mechanism}.
Several other mechanisms have been proposed, a general overview of
the research work on differential privacy can be found in
\cite{dwork2011firm, blocki2012johnson}.

\subsection{Differential Privacy in Social Networks}\label{sec:social}
Many efforts have been made towards publishing differential private
graph data. A work presented in \cite{sala2011sharing} seeks a
solution to share meaningful graph datasets, based on $dk-graph$
model, while preserving differential privacy. Another work in
preserving the degree distribution of a social network graph is
presented in \cite{hay2009accurate}. In
\cite{mir2012differentially}, differential privacy on a graph is
guaranteed by perturbing Kronecker model parameters. In
\cite{kenthapadi2012privacy}, the authors developed a differential
private algorithm that preserves distance between any two samples in
a given database. Although these studies deal with differential
private publication of social network data, none of them address the
utility of preserving the eigenvectors of the graph, the central
theme of this work.

Recently, several algorithms were proposed, mostly in theoretical
community, for publishing a differential private copy of the data
that preserves the top eigenvectors of the original dataset.
In~\cite{blum2005practical}, the authors propose to publish the
covariance matrix of the original data contaminated by random noise.
In~\cite{proserpio2012workflow, blocki2012johnson}, the authors show
that random projection by itself can preserve both the differential
privacy and the eigen spectrum of a given matrix provided
appropriate modification is made to the original matrix.
In~\cite{proserpio2012workflow}, the authors also present a
randomized response approach which achieves the preservation of
differential privacy and top eigenvectors by inverting each feature
attribute with a fixed probability. The main drawback of applying
these approaches to social network analysis is their high demand in
both computation and storage space. In particular, all these
approaches require, either explicitly or implicitly, generating a
large dense matrix of size $n\times n$, where $n$ is the number of
users in the network. For a social network of $10$ million users,
they need to manipulate a matrix of size $10^{14}$, which requires a
storage space of a few petabyes. In contrast, for the same social
network, if we assume most users have no more than $100$ links, the
graph of social network can be represented by a {\it sparse} matrix
that consumes only several gigabytes memory.

%

Besides publishing a differential private copy of data, an alternative approach is to publish differential privacy preserved eigenvectors. In~\cite{wangdifferential}, the authors propose to publish eigenvectors perturbed by Laplacian random noise, which unfortunately requires a large amount of random perturbation for differential privacy preservation and consequentially leads to a poor utility of data. An iterative algorithm was proposed in~\cite{hardt2012beyond} to compute differential private eigenvectors. It generates large dense matrix of $n\times n$ at each iteration, making it unsuitable for large-scale social network analysis. Sampling approaches based on the exponential mechanism are proposed in~\cite{chaudhuri2012near,kapralov2012differentially} for computing differential private singular vectors. Since these approaches require sampling very high dimensional vectors from a random distribution, they are computationally infeasible for large social networks.

\section{Differential Private Publication of Social Network Graph by Random Matrix}
\label{sec:pre}

In this section, we first present the proposed approach for
differential private publication of social network graph based on the random matrix
theory. We then present its guarantee on differential privacy and
the approximation of eigenvectors.

Let $G$ be a binary graph representing the connectivity of a social
network, and let $A \in \{0, 1\}^{n\times n}$ be the adjacency
matrix representing the graph, where $A_{i,j}=1$ if there is an edge
between nodes $i$ and $j$, and $A_{i,j}=0$, otherwise. By assuming
that the graph is undirected, $A$ will be a symmetric matrix, i.e.
$A_{i,j} = A_{j, i}$ for any $i$ and $j$. The first step of our
approach is to generate two Gaussian random matrix $P \in \mathbb
 R^{n\times m}$ and $Q \in \mathbb R^{m\times m}$, where $m \ll n$ is
the number of random projections. Here, each entry of $P$ is sampled
independently from a Gaussian distribution $\mathcal{N}(0, 1/m)$,
and each entry of $Q$ is sampled independently from another Gaussian
distribution $\mathcal{N}(0, \sigma^2)$, where the value of $\sigma$
will be discussed later. Using Gaussian random matrix $P$, we
compute the projection matrix $A_p \in \mathbb R^{n\times m}$ by $A_p = A\times P$, which
projects each row of $A$ from a high dimensional space $\R^n$ to into a low dimensional space $\R^m$.
We then perturb $A_p$ with the Gaussian random matrix $Q$ by
$\widehat{A} = A_p + Q$, and publish $\widehat{A}$ to the external world. Algorithm~\ref{alg:0} highlights the key
steps of the proposed routine for publishing the social network
graph. Compared to the existing approaches for differential private publication of social network graphs, the proposed algorithm is advantageous in three aspects:
\begin{itemize}
\item The proposed algorithm is computationally efficient as it does not require either storing or manipulating a {\it dense} matrix of $n\times n$.
\item The random projection matrix $P$ allows us to preserve the top eigenvectors of $A$ due to the theory of random matrix.
\item It is the {\it joint} effort between the random projection $P$ and the random perturbation $Q$ that leads to the preservation of differential privacy. This unique feature allows us to introduce a small amount of random perturbation for differential privacy preservation, thus improving the utility of data.
\end{itemize}

\begin{algorithm}
\small \LinesNumbered \DontPrintSemicolon \KwIn{(1) symmetric adjacency
matrix $A\in \mathbb R^{n\times n}$\\
\hspace{0.39in}(2) the number of random projections $m<n$\\
\hspace{0.39in}(3) variance for random noise $\sigma^2$\\
} \KwOut{$\widehat{A}$}

\BlankLine Compute a random projection matrix $P$, with $P_{i,j}
\sim \mathcal{N}(0, 1/m)$\; Compute a random perturbation matrix
$Q$, with $Q_{i,j} \sim \mathcal{N}(0, \sigma^2)$\; Compute the
projected matrix $A_p = A P$\; Compute the randomly perturbed matrix
$\widehat{A} = A_p + Q$\; \caption{$\widehat{A} =
\texttt{Publish}(A, m, \sigma^2)$}\label{alg:0}
\end{algorithm}

\subsection{Theoretical Analysis}
\label{sec:theory} In this section we give a theoretical analysis of
two main aspects of publishing differential private graph of social
networks. First, we prove theoretically that using random matrix for
publishing social network graphs guarantees differential privacy.
Next we give theoretical error bounds for approximating top$-k$
eigenvectors.
\subsubsection{Theoretical Guarantee on Differential Privacy}
\label{sec:diff}

Before we show the guarantee on differential privacy, we first
introduce the definition of differential privacy.

\begin{defn}
$(\epsilon, \delta)$-Differential Privacy: A (randomized) algorithm
$\mathcal{A}$ satisfies $(\epsilon, \delta)$-differential privacy,
if for all inputs $X$ and $X_0$ differing in at most one user's one
attribute value, and for all sets of possible outputs $D \subseteq
Range(\mathcal{A})$, we have
\begin{eqnarray}
\Pr\left(\mathcal{A}(X) \in D\right) \leq e^{\epsilon}
\Pr\left(\mathcal{A}(X_0) \in D \right) + \delta, \label{eqn:dp}
\end{eqnarray}
where the probability is computed over the random coin tosses of the
algorithm.
\end{defn}

To understand the implication of $(\epsilon, \delta)$-differential
privacy, consider the database $X \in \{0, 1\}^{n\times m}$ as a
binary matrix. Let $p_{i,j} : = \Pr(X_{i,j} = 1)$ represent the
prior knowledge of an attacker about $X$, and let $p'_{i,j} =
\Pr(X_{i,j} = 1|\mathcal{A}(X))$ represent his knowledge about $X$
after observing the output $\mathcal{A}(X)$ from algorithm
$\mathcal{A}$. Then, if an algorithm $\mathcal{A}$ satisfies
$(\epsilon, \delta)$-differential privacy, then with a probability
$1 - \delta$, we have, for any $i \in [n]$ and $j \in [m]$
\[
\left|\ln p_{i,j} - \ln p'_{i,j}\right| \leq \epsilon
\]
In other words, the additional information gained by observing
$\mathcal{A}(X)$ is bounded by $\epsilon$. Thus, parameter $\epsilon
> 0$ determines the degree of differential privacy: the smaller the
$\epsilon$, the less the amount of information will be revealed.
Parameter $\delta \in (0, 1)$ is introduced to account the rare
events when the two probabilities $\Pr\left(\mathcal{A}(X) \in
D\right)$ and $\Pr\left(\mathcal{A}(X_0) \in D \right)$ may differ
significantly from each other.

\begin{thm} \label{thm:dp}
Assuming $\delta < 1/2$, $n \geq 2$, and
\[
\sigma \geq \frac{1}{\epsilon}\sqrt{10\left(\epsilon + \ln\frac{1}{2\delta}\right)\ln\frac{n}{\delta}}
\]
Then, Algorithm~\ref{alg:0} satisfies $(\epsilon,
\delta)$-differential privacy w.r.t. a change in an individual
person's attribute.
\end{thm}
The detailed proof of Theorem~\ref{thm:dp} can be found in Section
\ref{sec:dpProof}. The key feature of Theorem~\ref{thm:dp} is that
the variance for generating the random perturbation matrix $Q$ is
$O(\ln n)$, almost independent from the size of social network. As a
result, we can ensure differential privacy for the published
$\widehat{A}$ for a very large social network by only introducing a
Gaussian noise with small variance, an important feature that allows
us to simultaneously preserve both the utility and differential
privacy. Our definition of differential privacy is a generalized
version of $\epsilon$-differential privacy which can be viewed as
$(\epsilon,0)$-differential privacy.

\subsubsection{Theoretical Guarantee on Eigenvector Approximation}
\label{sec:tranform} Let $u_1, \ldots, u_n$ the eigenvectors of the
adjacency matrix $A$ ranked in the descending order of eigenvalues
$\lambda_1, \ldots, \lambda_n$. Let $k$ be the number of top
eigenvectors of interests. Let $\ut_1, \ldots, \ut_k$ be the first
$k$ eigenvectors of $\widehat{A}$. Define the approximation error
for the first $k$ eigenvectors as
\[
\Er^2 = \max\limits_{1 \leq i \leq k} |\u_i - \ut_i|^2
\]
Our goal is to show that the approximation error $\Er^2$ will be
small when the number of random projections $m$ is sufficiently
large.
\begin{thm} \label{thm:eig}
Assume (i) $m \geq c(k + k\ln k)$, where $c$ is
an universal constant given in~\cite{Sarlos:2006:Improved}, (ii) $n \geq
4(m+1)\ln(12m)$ and (iii) $\lambda_k - \lambda_{k+1} \geq
2\sigma\sqrt{2n}$. Then, with a probability at least $1/2$, we have
\[
\Er^2 \leq \frac{16\sigma^2 n}{(\lambda_k - \lambda_{k+1})^2} +
\frac{32}{\lambda_k^2}\sum_{i=k+1}^n\lambda_i^2
\]
\end{thm}
The corollary below simplifies the result in Theorem~\ref{thm:eig}
by assuming that $\lambda_k$ is significantly larger than the
eigenvalues $\lambda_{k+1}, \ldots, \lambda_n$.
\begin{corollary}\label{cor:1}
Assume (i) $\lambda_k = \Theta(n/k)$, and (ii)
$\sum_{i=k+1}^n\lambda_i^2 = O(n)$. Under the same assumption for
$m$ and $n$ as Theorem~\ref{thm:eig}, we have, with a probability at
least $1/2$,
\[
\Er \leq O\left(k\left[\frac{\sigma}{\sqrt{n}} + \frac{1}{\sqrt{n}}\right]\right)
\]
\end{corollary}
As indicated by Theorem~\ref{thm:eig} and Corollary~\ref{cor:1},
under the assumptions (i) $\lambda_k$ is significantly larger than
eigenvalues $\lambda_{k+1}, \ldots, \lambda_n$, (ii)
the number of random projections $m$ is sufficiently larger than
$k$, and (iii) $n$ is significantly larger than the number of random
projections $m$, we will have the approximation error $\Er \propto
O(k/\sqrt{n})$ in recovering the eigenvectors of the adjacency
matrix $A$. We also note that according to Corollary~\ref{cor:1},
the approximation error is proportional to $\sigma$, which
measures the amount of random perturbation needed for differential
privacy preservation. This is consistent with our intuition, i.e.
the smaller the random perturbation, the more accurate the
approximation of eigenvectors.

\subsubsection{Proof of Theorem~\ref{thm:dp}}\label{sec:dpProof} To prove that
Algorithm~\ref{alg:0} is differential private, we need the following
theorem from~\cite{kenthapadi2012privacy}
\begin{lemma} (Theorem 1~\cite{kenthapadi2012privacy})
Define the $\ell_2$-sensitivity of the projection matrix P as
$w_2(P) = \max\limits_{1 \leq i \leq n} |P_{i,*}|_2$, where
$P_{i,*}$ represents the $i$th row of matrix $P$. Assuming $\delta <
1/2$, and
\[
\sigma \geq \frac{w_2(P)}{\epsilon}\sqrt{2\left(\epsilon +
\ln\frac{1}{2\delta} \right)}
\]
Then Algorithm~\ref{alg:0} satisfies $(\epsilon,
\delta)$-differential privacy w.r.t. a change in an individual
person's attribute.
\end{lemma}
In order to bound $w_2(P)$, we rely on the following concentration
for $\chi^2$ distribution.
\begin{lemma} (Tail bounds for the $\chi^2$ distribution ) Let $X_1, \ldots, X_d$ be independent draws from $\N(0, 1)$. Therefore, for any $0 < \delta < 1$, we have, with a probability $1 - \delta$,
\[
\sum_{i=1}^d X_i^2 \leq d + 2\sqrt{d\ln\frac{1}{\delta}} +
2\ln\frac{1}{\delta}
\]
\end{lemma}
Define
\[
    z^2_i = \sum_{j=1}^m P_{i,j}^2
\]
Evidently, according to the definition of $w^2_2(P)$, we have
\[
    w^2_2(P) = \max\limits_{1 \leq i \leq n} z_i^2
\]
Since $P_{i,j} \sim \N(0, 1/m)$, we have $mz_i^2$ follow the
$\chi^2$ distribution of $d$ freedom. Using Lemma 2, we have, with a
probability $1 - \delta$,
\[
    z^2_i \leq 1 + 2\sqrt{\frac{1}{m}\ln\frac{1}{\delta}} + \frac{2}{m}\ln\frac{1}{\delta}
\]
By taking the union bound, we have, with a probability $1 - \delta$
\begin{eqnarray}
w^2_2(P) = \max\limits_{1 \leq i \leq m} z_i^2\leq 1 +
2\sqrt{\frac{1}{m}\ln\frac{n}{\delta}} +
\frac{2}{m}\ln\frac{n}{\delta} \leq 2 \label{eqn:temp-1}
\end{eqnarray}
where the last inequality follows from $m \geq 4\ln(n/\delta)$. We
complete the proof by combining the result from Lemma 1 and the
inequality in (\ref{eqn:temp-1}).

\subsubsection{Proof of Theorem~\ref{thm:eig}}

Let $A \in \R^{n\times n}$ be the adjacency matrix, $A_p = AP$, and
$\Ah = A_p + Q$. Let $\uh_1, \ldots, \uh_k$ be the first $k$
eigenvectors of matrix $A_p$. Define $U = (\u_1, \ldots, \u_k)$,
$\Uh = (\uh_1, \ldots, \uh_k)$, and $\Ut = (\ut_1, \ldots, \ut_k)$.
For each of these matrices, we define a projection operator, denoted
by $P_k$,  $\Ph_k$ and $\Pt_k$, as
\begin{eqnarray*}
P_k & = & \sum_{i=1}^k \u_i \u_i^{\top} = UU^{\top} \\
\Ph_k & = & \sum_{i=1}^k \uh_i\uh_i^{\top} = \Uh\Uh^{\top} \\
\Pt_k & = & \sum_{i=1}^k \ut_i\ut_i^{\top} = \Ut\Ut^{\top}
\end{eqnarray*}
We first bound the approximation error $\Er^2$ by the difference
between projection operators, i.e.
\[
\Er^2 = \max\limits_{1 \leq i \leq k} |\u_i - \ut_i|^2 \leq
\|UU^{\top} - \Ut\Ut^{\top}\|_2 = \|P_k - \Pt_k\|_2
\]
where $\|\cdot\|_2$ stands for the spectral norm of matrix. Using
the fact that
\begin{eqnarray}
\Er^2 & \leq & \|P_k - \Pt_k\|_2^2 = \|P_k - \Ph_k + \Ph_k - \Pt_k\|_2^2 \nonumber \\
& \leq & 2\|P_k - \Ph_k\|_2^2 + 2\|P_k - \Pt_k\|_2^2 \nonumber \\
& \leq & 2\|P_k - \Ph_k\|_F^2 + 2\|P_k - \Pt_k\|_2^2
\label{eqn:bound-3}
\end{eqnarray}
where $\|\cdot\|_F$ stands for the Frobenius norm of matrix, below
we will bound $\|P_k - \Ph_k\|_F$ and $\|P_k - \Pt_k\|_F$,
separately.

To bound $\|P_k - \Ph_k\|_F$, we need the following theorem for
random matrix.
\begin{lemma} (Theorem 14~\cite{Sarlos:2006:Improved}) Assume $0 < \epsilon \leq 1$ and $m \geq c(k/\epsilon + k\ln k)$, where $c$ is some universal constant. Then, with a probability at least $2/3$, we have
\[
    \|A - \Ph_k(A)\|_F \leq (1 + \epsilon)\|A - P_k(A)\|_F,
\]
\end{lemma}
Since
\begin{eqnarray*}
\lefteqn{\scriptstyle\|A - \Ph_k(A)\|_F \geq - \|A - P_k(A)\|_F + \|P_k(A) - \Ph_k(A)\|_F} \\
& = & \scriptstyle - \|A - P_k(A)\|_F - \|P_k(A) + \Ph_kP_k(A) + \Ph_kP_k(A) - \Ph_k(A)\|_F \\
& \geq &\scriptstyle - \|A - P_k(A)\|_F + \|P_k(A) - \Ph_kP_k(A)\|_F - |\Ph_k(A - P_k(A))\|_F\\
& \geq &\scriptstyle \|P_k(A) - \Ph_kP_k(A)\| - 2\|A - P_k(A)\|_F>,
\end{eqnarray*}
combining with the result from Lemma 3, we have, with a probability
at least $2/3$,
\begin{eqnarray}
\|(P_k - \Ph_kP_k)(A)\|_F \leq (3+\epsilon)|A - P_k(A)|_F
\label{eqn:temp-2}
\end{eqnarray}
Since
\begin{eqnarray*}
& & \|(P_k - \Ph_kP_k)(A)\|_F \\
& = &\|(P_kP_k - \Ph_kP_k)(A)\|_F=\|(P_k - \Ph_k)P_k(A)\|_F \\
& \geq & \|P_k - \Ph_k\|_F\|P_k(A)\|_2=\lambda_k\|P_k - \Ph_k\|_F
\end{eqnarray*}
combining with the inequality in (\ref{eqn:temp-2}), we have, with a
probability at least $2/3$,
\begin{eqnarray}
\|P_k - \Ph_k\|_F \leq \frac{3 + \epsilon}{\lambda_k}|A - P_k(A)|_F
\label{eqn:bound-2}
\end{eqnarray}

In order to bound $\|\Ph_k - \Pt_k\|_2$, we use the Davis-Kahan
sin$\Theta$ theorem given as below.
\begin{lemma}
Let $A$ and $\tilde{A}$ be two symmetric matrices. Let
$\{\u_i\}_{i=1}^k$ and $\{\ut_i\}_{i=1}^k$ be the first $k$
eigenvectors of $A$ and $\tilde{A}$, respectively. Let
$\lambda_k(A)$ denote the $k$th eigenvalue of $A$. Then, we have
\[
\|P_k - \Pt_k\|_2 \leq \frac{\|A - \tilde{A}\|_2}{\lambda_k(A) -
\lambda_{k+1}(\tilde{A})}
\]
if $\lambda_k(A) > \lambda_{k+1}(\tilde{A})$, where $P_k =
\sum_{i=1}^k \u_k\u_k^{\top}$ and $\Pt_k = \sum_{i=1}^k
\ut_i\ut_i^{\top}$.
\end{lemma}
Using Lemma 4 and the fact
\[
\lambda_{k+1}(\Ah) \leq \lambda_{k+1}(A_p) + \|A_p - \Ah\|_2 =
\lambda_k + \|Q\|_2
\]
we have
\begin{eqnarray*}
\|\Ph_k - \Pt_k\|_2 & \leq & \frac{\|A_p - \Ah\|_2}{\lambda_k(A_p) - \lambda_{k+1}(\Ah)} \\
& \leq & \frac{\|Q\|_2}{\lambda_k - \lambda_{k+1} - \|Q\|_2}
\end{eqnarray*}
Under the assumption that $\lambda_k - \lambda_{k+1} \geq 2\|Q\|_2$,
we have
\begin{eqnarray*}
\|\Ph_k - \Pt_k\|_2 \leq \frac{2\|Q\|_2}{\lambda_k - \lambda_{k+1}}
\end{eqnarray*}
In order to bound the spectral norm of $Q$, we need the following
lemma from random matrix.
\begin{lemma}
Let $A \in \R^{r\times m}$ be a standard Gaussian random matrix. For
any $0 <\epsilon \leq 1/2$, with a probability at least $1 -
\delta$, we have
\[
    \left\|\frac{1}{m}AA^{\top} - I\right\|_2 \leq \epsilon
\]
provided
\[
m \geq \frac{4(r+1)}{\epsilon^2}\ln\frac{2r}{\delta}
\]
\end{lemma}
Using Lemma 5 and the fact that $Q_{i,j} \sim \N(0, \sigma^2)$, we
have, with a probability at least $5/6$
\[
\|QQ^{\top}\|_2 \leq (1 + \eta)\sigma^2 n
\]
where
\[
n \geq \frac{4(m+1)}{\eta^2}\ln (12m)
\]
As a result, we have, with a probability at least $5/6$,
\[
\|Q\|_2 \leq \sigma\sqrt{(1 + \eta)n}
\]
and therefore
\begin{eqnarray}
\|\Ph_k - \Pt_k\|_2 \leq \frac{2\sigma}{\lambda_k -
\lambda_{k+1}}\sqrt{(1 + \eta)n} \label{eqn:bound-1}
\end{eqnarray}
We complete the proof by combining the bounds for $\|P_k -
\Ph_k\|_F$ and $\|\Ph_k - \Pt_k\|_2$ in (\ref{eqn:bound-2}) and
(\ref{eqn:bound-1}) and plugging them into the inequality in
(\ref{eqn:bound-3}).
%

\section{Experimental Results}\label{sec:results}
To demonstrate the effectiveness of our differential private random
matrix approach and to illustrate the utility preservation of
eigen-spectrum, we perform experiments over graphs obtained from
three different online social networks.
We analyze the impact of perturbation by evaluating the utility of
the published data for two different applications which require
spectral information of a graph.
First, we consider clustering of social networks, which has been
widely used for community detection in social networks. We choose
spectral clustering algorithm in our study, which depends on the
eigenvectors of the adjacency matrix. Next, we examine the problem
of identifying the ranks of influential nodes in a social network
graph.

For the evaluation purposes, we obtain clusters and node ranks from
the published graph, and compare the results against those obtained
from the original graph. We give a brief description of the results
obtained for each of the applications of graph spectra in the
subsequent sections.

\subsection{Dataset} \label{sec:data}
In our evaluation we use three different social network graphs from
Fcaebook, Live Journal and Pokec.
We use the Facebook data set collected by Wilson et al. from
Facebook \cite{wilson2009user}.
The social graphs of Live Journal and Pokec were obtained from
publicly available SNAP graph library
\cite{takacdata},\cite{yang2012defining}.
The choice of these social networks is based on two main
requirements.
First, the network should be large enough so that it is a true
representation of real online social structure.
A small network not only under-represents the social structure, but
also produces biased results.
Second, the number of edges in the network should be sufficiently
large in order to reveal the interesting structure of the network.
For all three benchmark datasets, the ratio of the number of edges
to the number of nodes is between $7$ and $20$. Table
\ref{table:data} provides the basic statistics of the social network
graphs.

\begin{table}
\centering
\begin{tabular}{|c|c|c|}
  \hline
  Network & Nodes & Edges \\\hline
  Facebook & $3,097,165$ & $23,667,394$ \\\hline
  Pokec & $1,632,803$ & $30,622,564$  \\\hline
  LiveJournal & $3,997,962$ & $34,681,189$  \\
  \hline
\end{tabular}
\caption{Dataset Description} \label{table:data}
\end{table}

Figure \ref{fig:nodeDist} shows degree distribution of three online
social networks on log-log scale. We can see that the data follows a
power law distribution which is a characteristic of social network
degree distribution.

\begin{figure}[t]
\centering
\includegraphics[width=0.8\columnwidth]{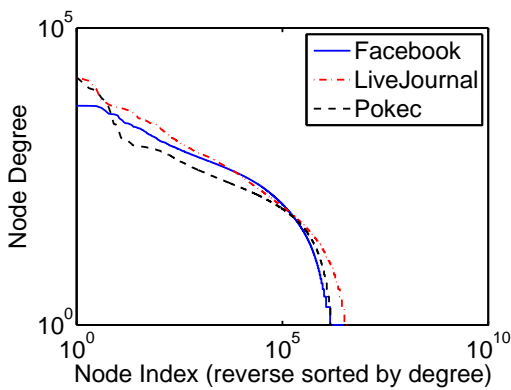}
\caption{Degree distribution of three datasets.}
\label{fig:nodeDist}
\end{figure}

\subsection{Spectral Clustering} \label{sec:spectral} Clustering is a
widely used technique for identifying groups of similar instances in
a data.
Clustering has applications in community detection, targeted
marketing, bioinformatics etc.
Social networks posses large amount of information which can be
utilized in extensive data mining applications.
Large complex graphs can be obtained from social networks which
represent relationships among individual users.
One of the key research questions is the understanding of community
structure present in large social network graphs.
Social networking platforms possess strong community structure of
users, which can be captured by clustering nodes of a social network
graph.
Detecting communities can help in identifying structural position of
nodes in a community.
Nodes with a central position in a community have influence in the
community.
Similarly, nodes lying at the intersection of two communities are
important for maintaining links between communities.
Disclosure of the identity of such nodes having important structural
properties results in serious privacy issues.
Therefore, in order to protect an individual's privacy it is crucial
for data publishers to provide rigorous privacy guarantees for the
data to be published.

In our experiments, we use spectral clustering for evaluating our
privacy-preserving random matrix approach.
Spectral clustering has many fundamental advantages over other
clustering algorithms \cite{von2007tutorial}. Unlike other clustering algorithms, spectral clustering is
particularly suitable for social networks, since it requires an
adjacency matrix as an input and not a feature representation of the
data.
For social network data graph $G$ represented by the binary
adjacency matrix$A$, spectral clustering techniques
\cite{von2007tutorial} utilize the eigen-spectrum of $A$ to perform
clustering. The basic idea is to view clustering as a graph
partition problem, and divide the graph into several disjoint
subgraphs by only removing the edges that connect nodes with small
similarities. Algorithm \ref{alg:spectral} gives the standard
clustering algorithm, and Algorithm \ref{alg:diffspectral} states
the key steps of differential private spectral clustering algorithm.
Algorithm~\ref{alg:diffspectral} differs from
Algorithm~\ref{alg:spectral} in that it calls the publish routine in
Algorithm~\ref{alg:0} to obtain a differential private matrix which
represents the structure of a social network.

\begin{algorithm}
\small
\LinesNumbered \DontPrintSemicolon \KwIn{(1) Adjacency
Matrix $A\in \mathbb R^{n\times n}$\\
\hspace{0.41in}(2) Number of clusters $k$\\
} \KwOut{Clusters $C_1,...,C_k $}

\BlankLine
Compute first $k$ eigenvectors $\u_1,..,\u_k$ of $A$\; Get matrix $U\in \mathbb R^{n\times k}$
where $ith$ column of $U$ is $\u_i$\; Obtain clusters by applying
$k-$means clustering on matrix $U$\;
\caption{\textbf{Spectral Clustering}}\label{alg:spectral}
\end{algorithm}

\begin{algorithm}
\small
\LinesNumbered \DontPrintSemicolon \KwIn{(1) adjacency
matrix $A\in \mathbb R^{n\times n}$\\
\hspace{0.41in}(2) number of clusters $k$\\
\hspace{0.41in}(3) the number of random projections $m < n$\\
\hspace{0.41in}(4) variance for random noise $\sigma^2$\\
} \KwOut{Clusters $C_1,...,C_k $}

\BlankLine {Compute a differential private matrix for social network
$A$ by $\widehat{A} = \texttt{Publish}(A, m, \sigma^2)$} \; Compute
first $k$ eigenvectors $\ut_1,..,\ut_k$ of $\widehat{A}$\; Get
matrix $U\in \mathbb R^{n\times k}$ where $ith$ column of $U$ is
$\ut_i$\; Obtain clusters by applying $k-$means clustering on matrix
$U$\; \caption{\textbf{Differential Private Spectral
Clustering}}\label{alg:diffspectral}
\end{algorithm}

In order to evaluate the utility of the published data for
clustering, we utilize normalize mutual information (NMI) as a
measure to evaluate the quality of clustering
\cite{guiacsu1977information}.
Although \textit{Purity} is a simpler evaluation measure, high
purity is easy to achieve for large number of clusters and cannot be
used to evaluate trade off between quality of clustering and number
of clusters.
NMI allows us to evaluate this tradeoff by normalizing mutual
information $I(\omega ;C)$ as described in Equation \ref{eqn:nmi}.
\begin{equation}
    NMI = \frac{I(\omega ;C)}{[H(\omega)+H(C)]/2} \label{eqn:nmi},
\end{equation}
where $H$ is entropy which measures the uniformity of the
distribution of nodes in a set of clusters, $\omega = {w_1,...,w_k}$
is a set of clusters and $C = {c_1,...,c_k}$ is a set of classes or
ground truth. NMI is bounded between $0$ and $1$, and the larger the NMI, the better the clustering performance is.

We perform extensive experiments over the datasets to evaluate our
approach.
We now give a stepwise explanation of our evaluation protocol.
Since we donot have ground truth about the communities in the
datasets, we employ an exhaustive approach to evaluate clustering
over the original data and generate the ground truth communities.
First, for a given value of $k$ we generate $5$ different sets of
clusters from Algorithm \ref{alg:spectral}, represented as $C_{i}$
for $i=1,..,5$.
Since spectral clustering employs $k-$means, each set $C_i$ can have
different cluster distributions.
Therefore, to evaluate the consistency in cluster distribution, NMI
values are obtained for $5\choose 2$ different pairs of sets
represented as $(C_i,C_j)$, where $i\neq j$ and average value is
reported.
Then, another $5$ cluster sets are obtained through Algorithm
\ref{alg:diffspectral}, represented as $\omega_i$ for $i=1,...,5$.
Finally, to evaluate cluster sets $\omega_i$, NMI values are
obtained using $C_{i}$ as the ground truth.
In this case NMI values are obtained for each pair $(\omega_i,C_j)
\forall i,j\in {1,...,5}$ and average value is reported.

Since one of the advantages of the proposed approach is its low
sensitivity towards noise, we evaluate the clustering results for
three different values of $\sigma$, where $\sigma = 0.1,0.5$ and
$1$. We note that these values of random noise were suggested
in~\cite{kapralov2012differentially}, based on which we build our
theoretical foundation. For each $\sigma$, we evaluate clustering
for two different number of random projections $m=20,200$.

Figure \ref{fig:fnmi}, \ref{fig:ljnmi} and \ref{fig:pnmi} shows NMI
values obtained for four different values of $k$, where symbol $O$
represents the NMI values obtained by using the original data.
It is not surprising to observe that the clustering quality deteriorates with increasing number of clusters. This is because the
larger the number of clusters, the more the challenging the problem
is.
Overall, we observe that $m = 200$ yields significantly better
clustering performance than $m = 20$. When the random perturbation
is small (i.e. $\sigma = 0.1$), our approach with $m=200$ random
projections yields similar clustering performance as spectral
clustering using the original data. This is consistent with our
theoretical result given in Theorem~\ref{thm:eig}, i.e. with
sufficiently large number of random projections, the approximation
error in recovering the eigenvectors of the original data can be as
small as $O(1/\sqrt{n})$. Finally, we observe that the clustering
performance declines with larger noise for random perturbation.
However, even with random noise as large as $\sigma = 1$, the
clustering performance using the differential private copy of the
social network graph still yield descent performance with
$\mbox{NMI} \geq 0.70$. This is again consistent with our
theoretical result: the approximation error of eigenvectors is
$O(\sigma/\sqrt{n})$, and therefore will be small as long as
$\sigma$ is significantly smaller than $\sqrt{n}$. Finally, Table
\ref{table:timesU} shows the memory required for the published data
matrix and the time required to compute the random projection query
over the graph matrix. It is not surprising to see that both the
memory requirement and running time increases significantly with
increasing number of random projections.

 \begin{figure*}[!t]
 \centering
 \subfigure[\scriptsize $\sigma=0.1$]{
 {\includegraphics[width=0.6\columnwidth]{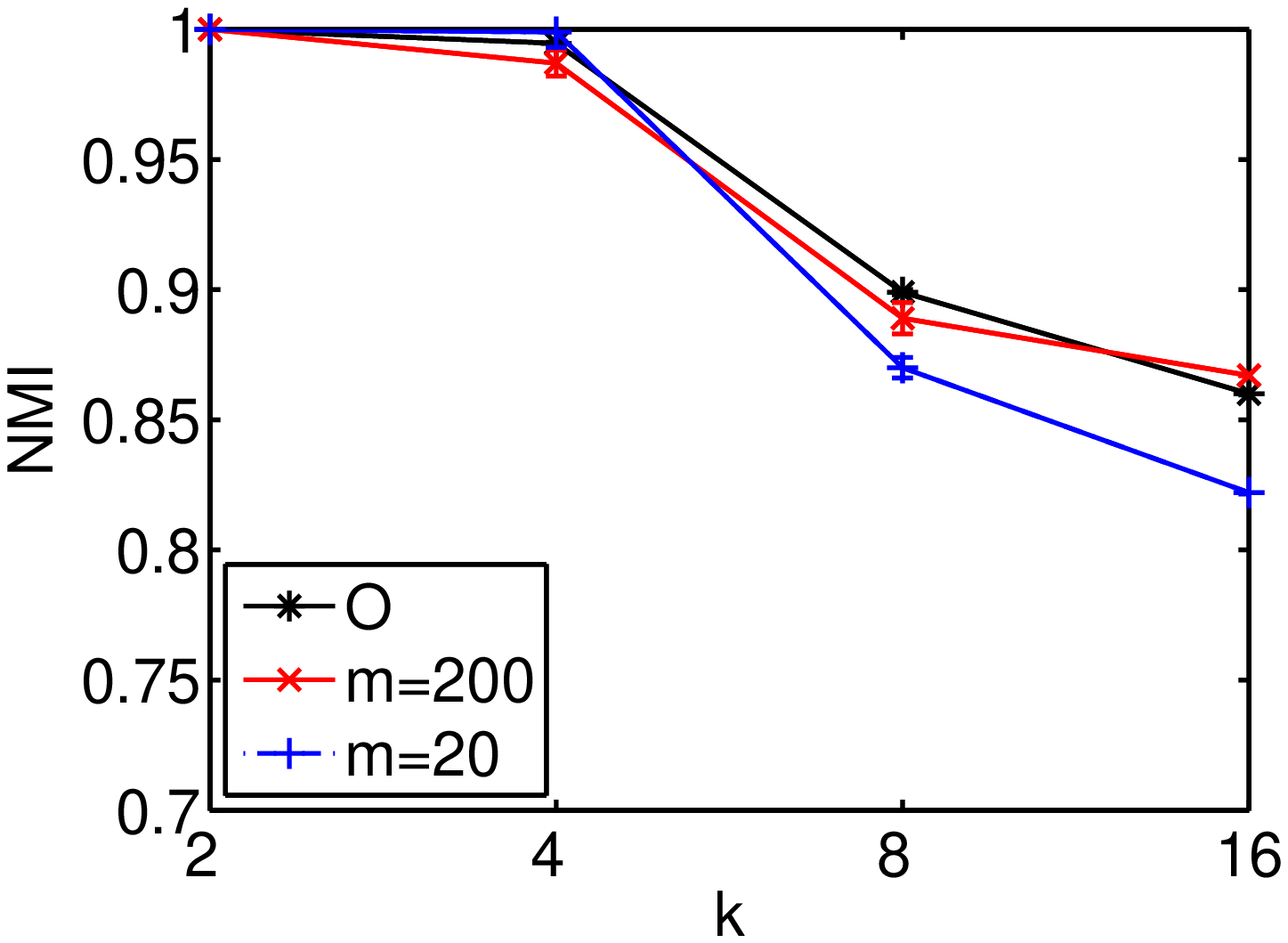}}
 \label{fig:fn01}}
 \hspace{-0.1in}
 \subfigure[\scriptsize $\sigma=0.5$]{
 {\includegraphics[width=0.6\columnwidth]{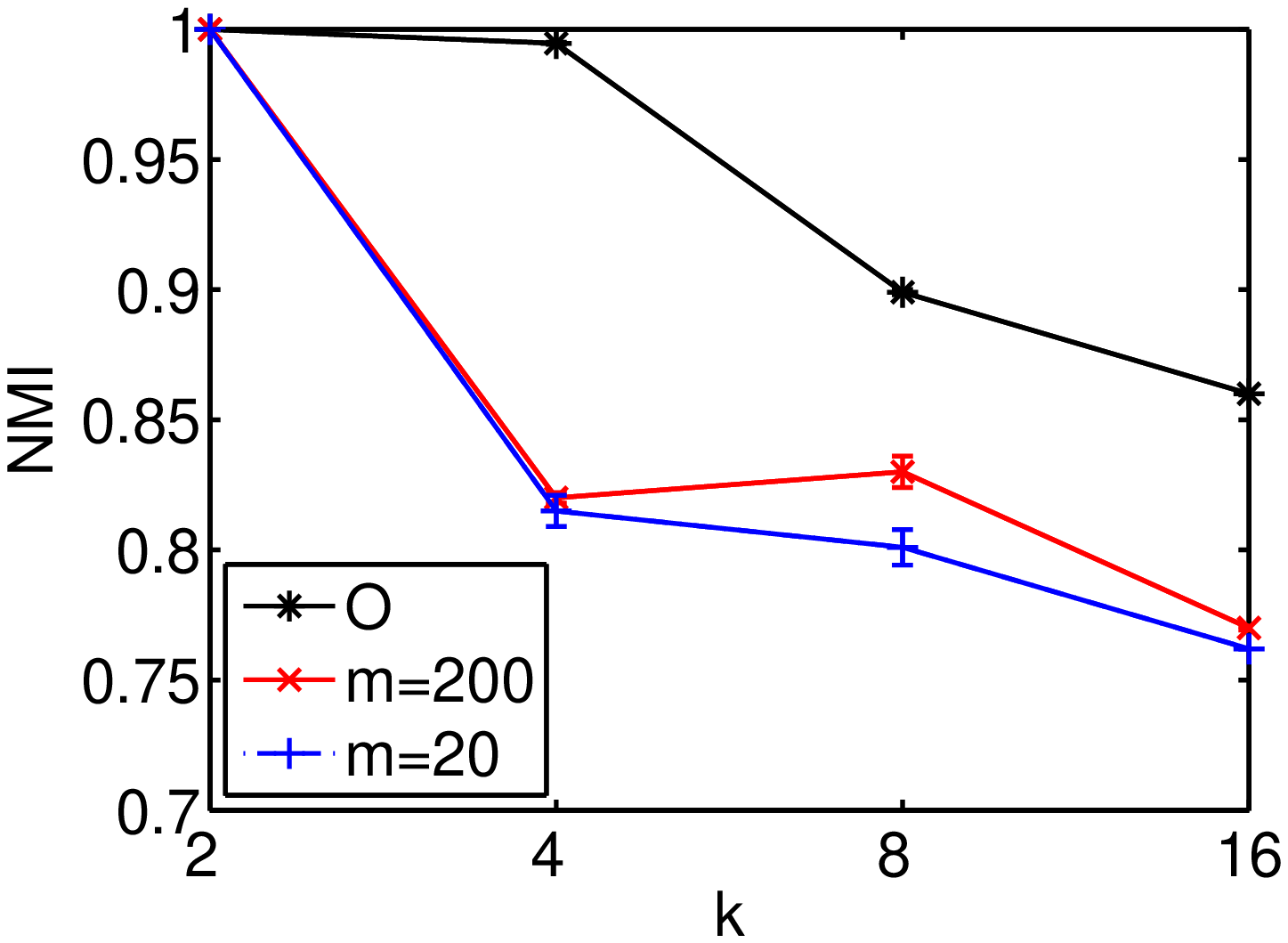}}
 \label{fig:fn05}}
 \hspace{-0.1in}
  \subfigure[\scriptsize $\sigma=1$]{
 {\includegraphics[width=0.6\columnwidth]{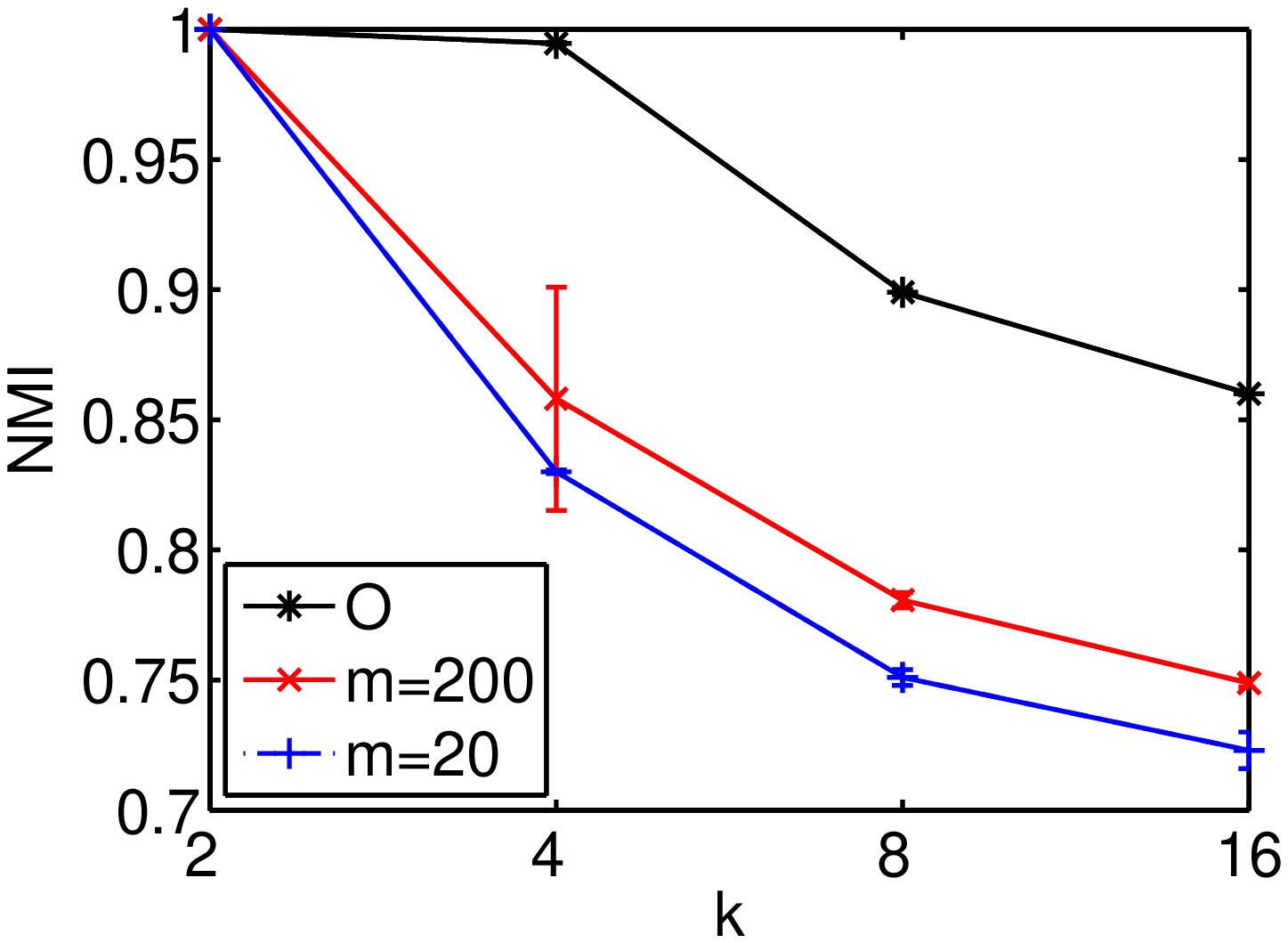}}
 \label{fig:fn1}}
 \caption{NMI values for Facebook}
 \label{fig:fnmi}
 \end{figure*}

 \begin{figure*}[!t]
 \centering
 \subfigure[\scriptsize $\sigma=0.1$]{
 {\includegraphics[width=0.6\columnwidth]{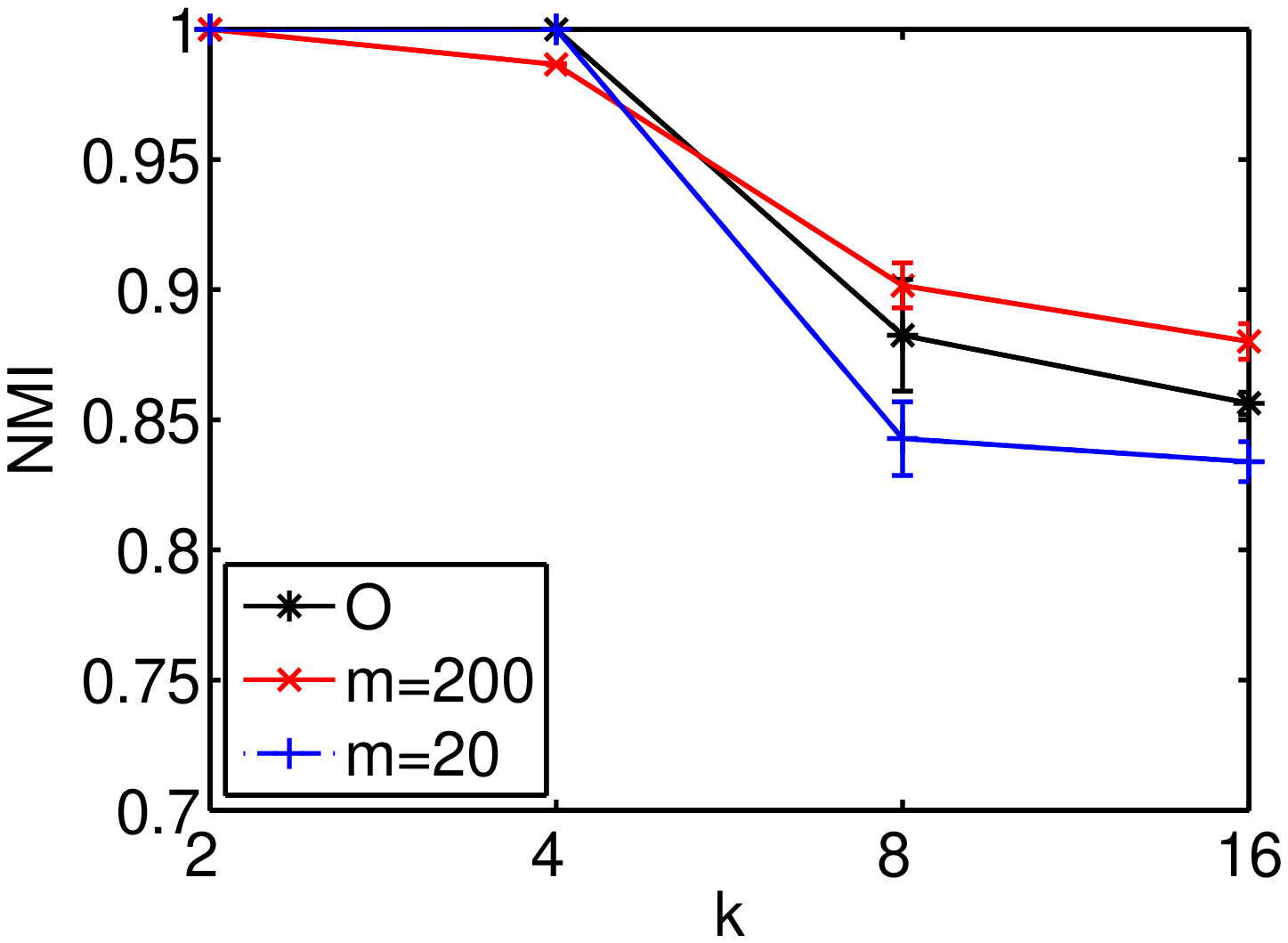}}
 \label{fig:lj01}}
 \hspace{-0.1in}
 \subfigure[\scriptsize $\sigma=0.5$]{
 {\includegraphics[width=0.6\columnwidth]{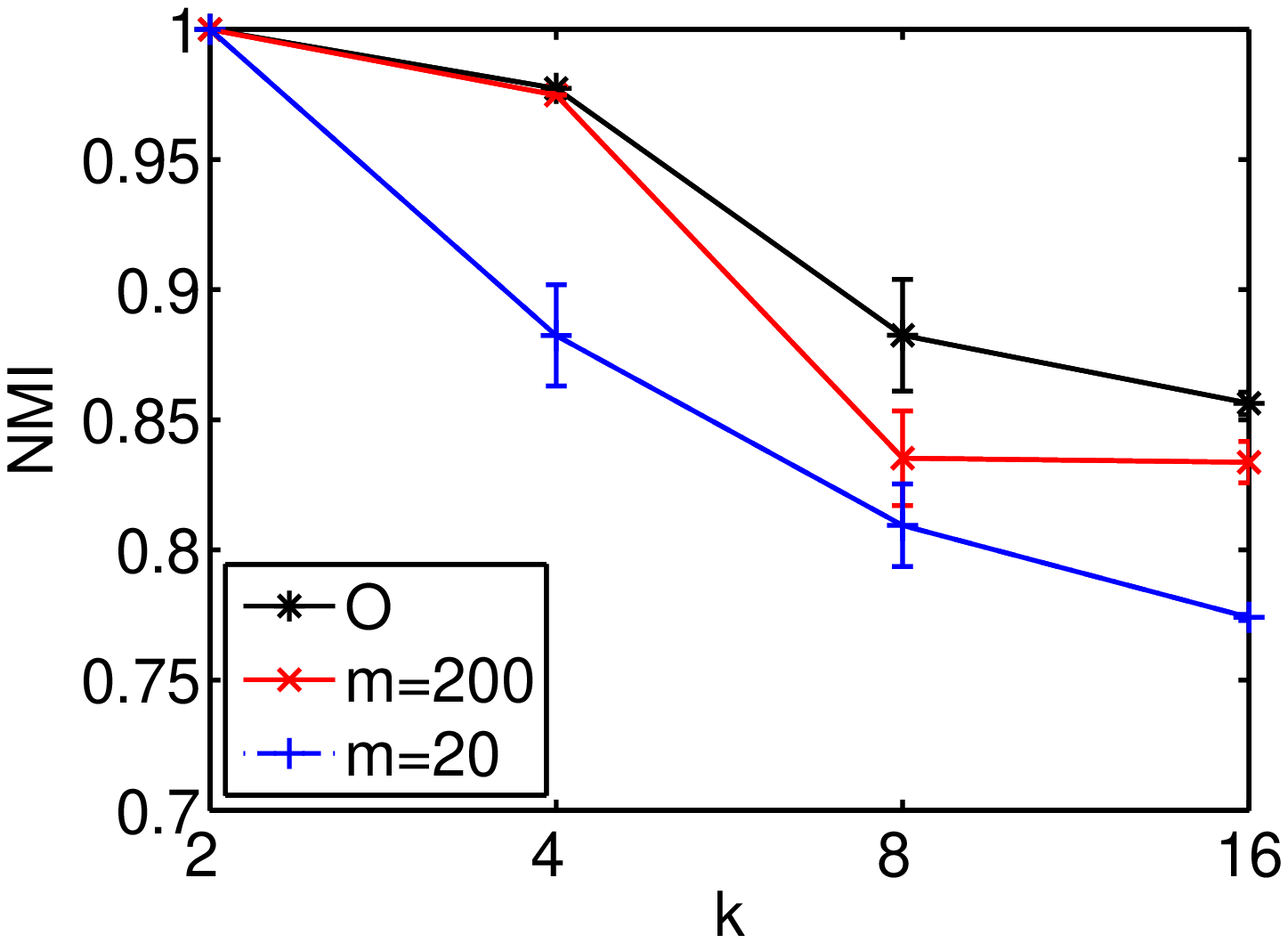}}
 \label{fig:lj05}}
 \hspace{-0.1in}
  \subfigure[\scriptsize $\sigma=1$]{
 {\includegraphics[width=0.6\columnwidth]{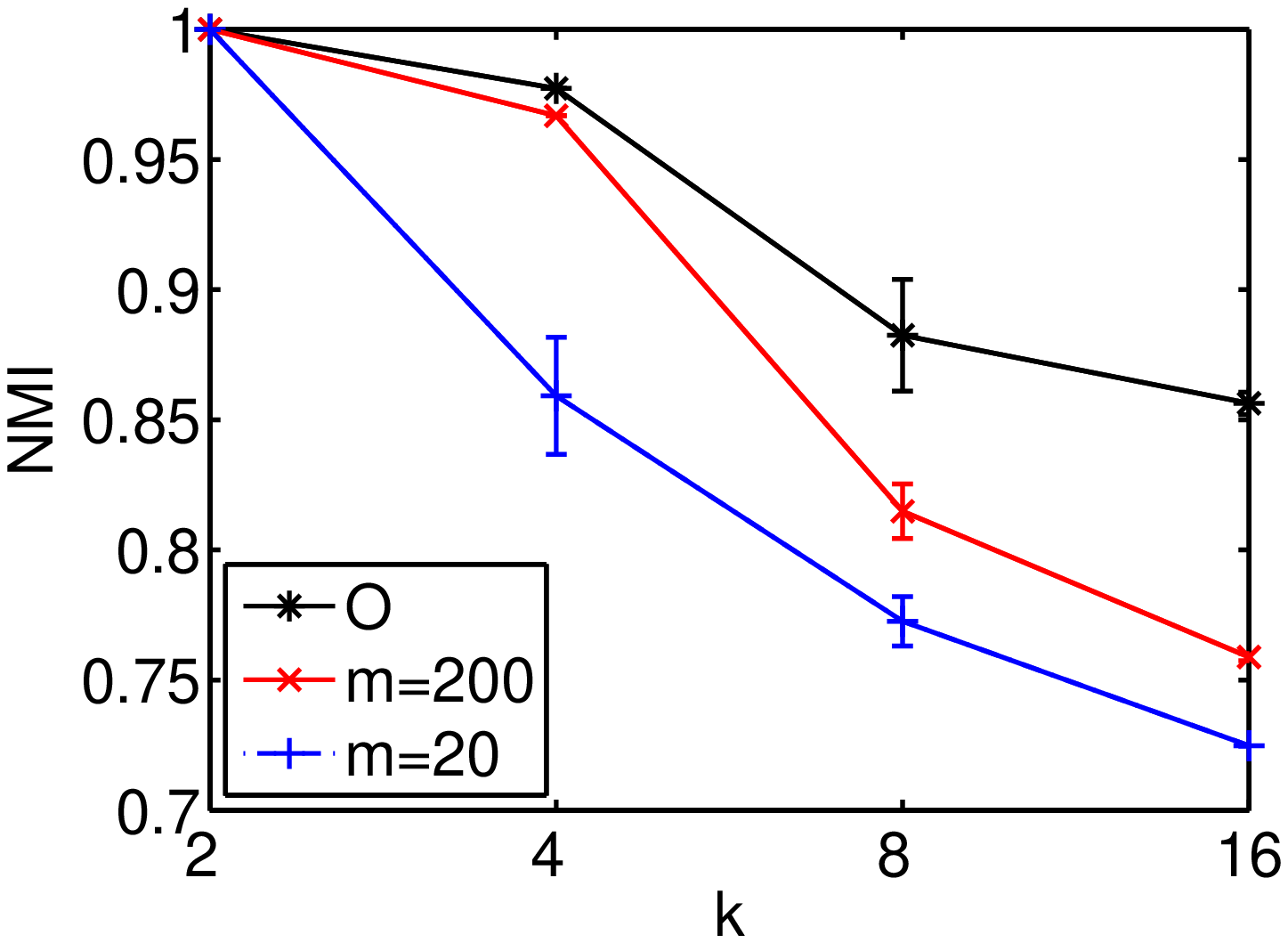}}
 \label{fig:lj1}}
 \caption{NMI values for Live Journal}
 \label{fig:ljnmi}
 \end{figure*}

 \begin{figure*}[!t]
 \centering
 \subfigure[\scriptsize $\sigma=0.1$]{
 {\includegraphics[width=0.6\columnwidth]{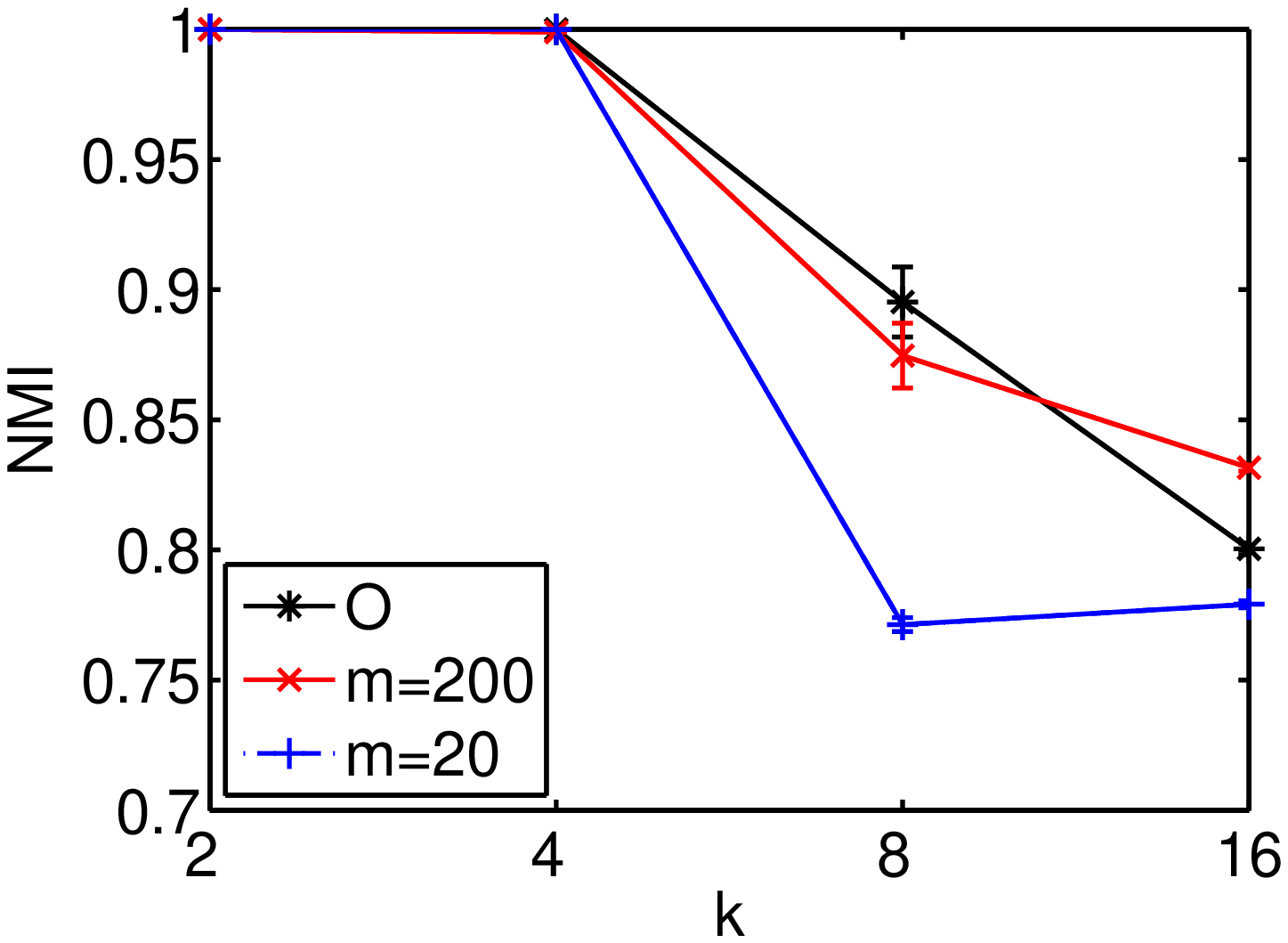}}
 \label{fig:pn01}}
 \hspace{-0.1in}
 \subfigure[\scriptsize $\sigma=0.5$]{
 {\includegraphics[width=0.6\columnwidth]{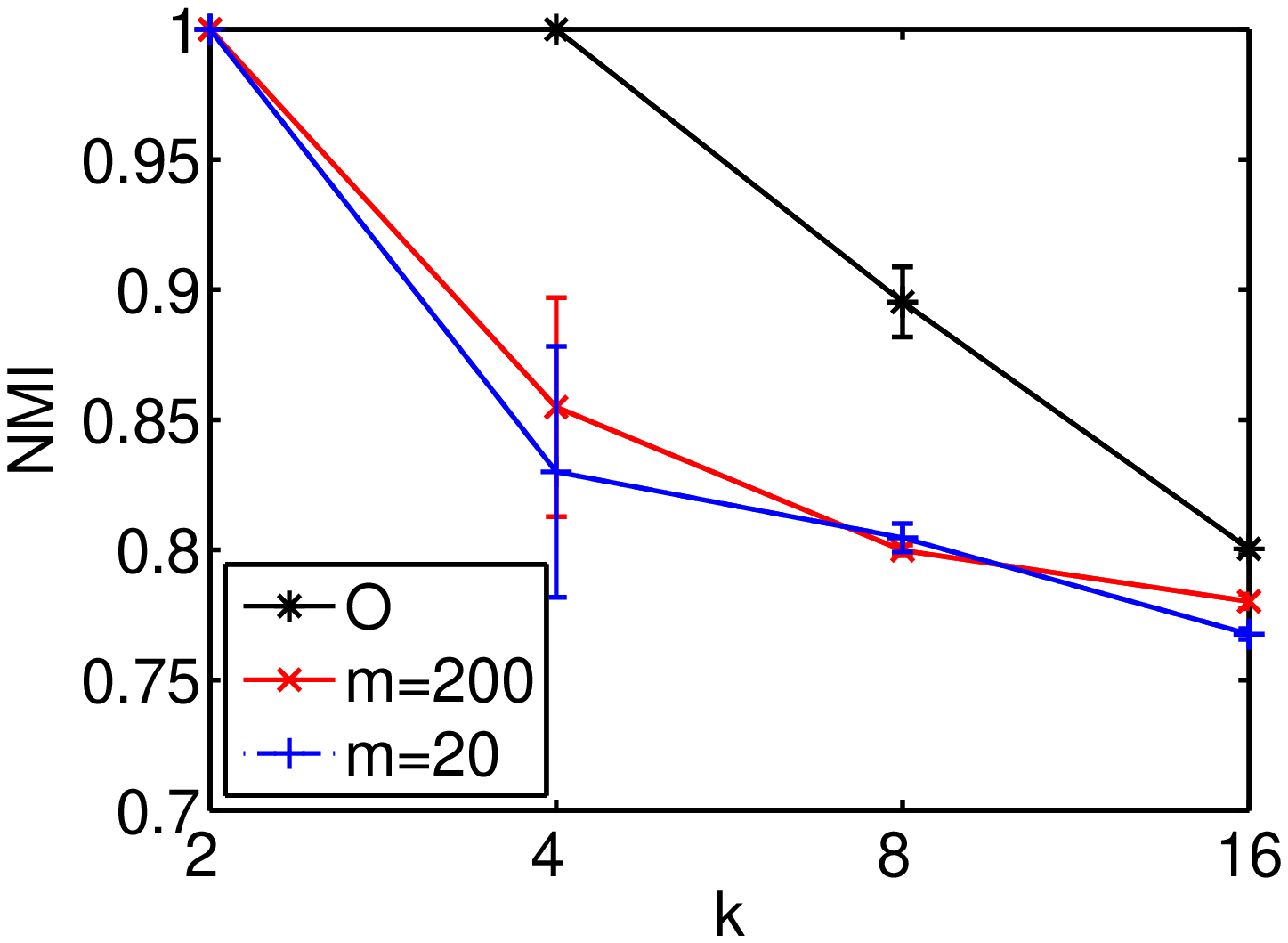}}
 \label{fig:pn05}}
 \hspace{-0.1in}
  \subfigure[\scriptsize $\sigma=1$]{
 {\includegraphics[width=0.6\columnwidth]{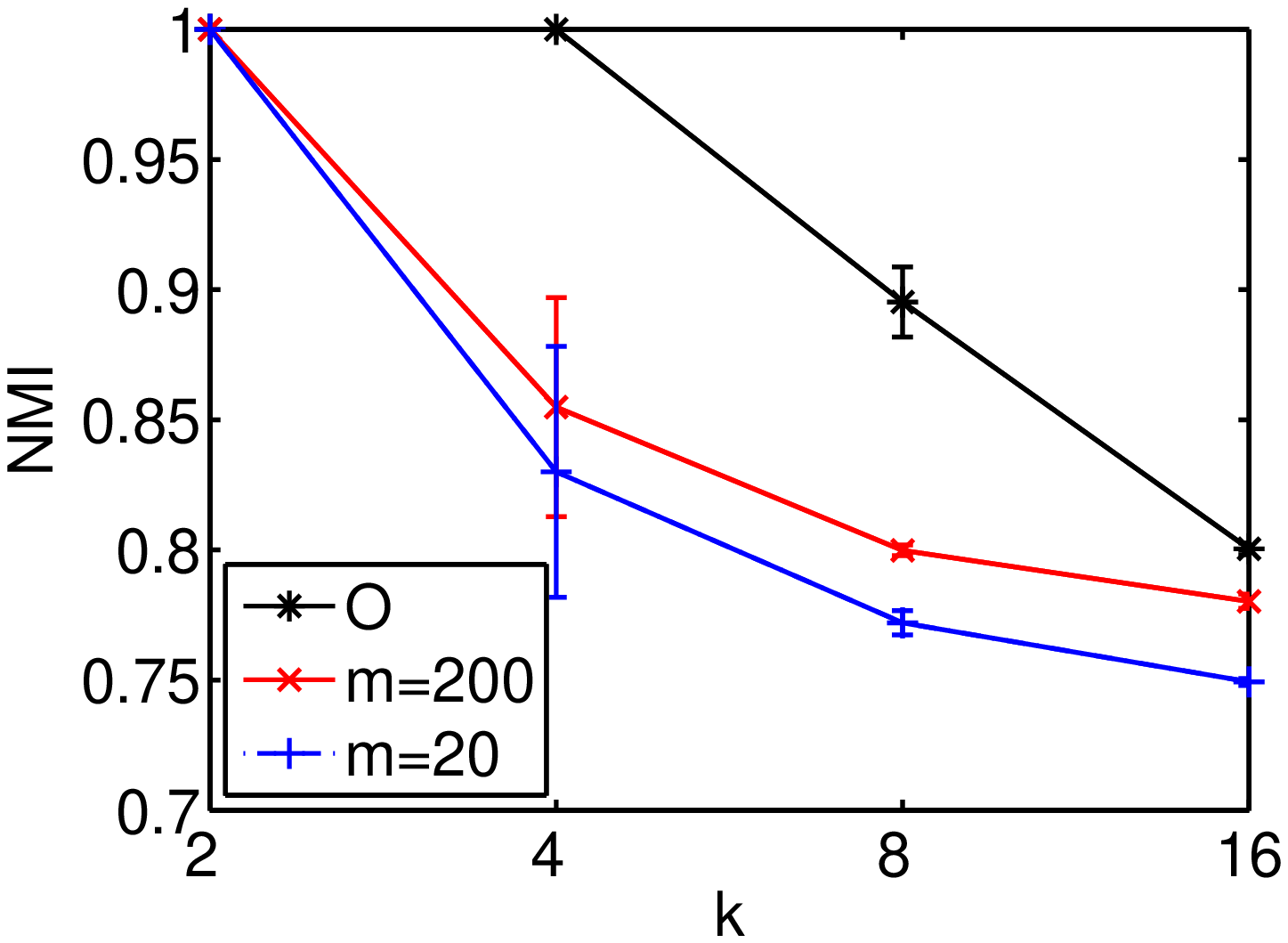}}
 \label{fig:pn1}}
 \caption{NMI values for Pokec}
 \label{fig:pnmi}
 \end{figure*}

To show the variation in the cluster distribution, we select
clusters obtained from Facebook data for $k=200$ and $\sigma=1$.
Figure \ref{fig:facebookC},\ref{fig:journalC} and \ref{fig:pokecC}
shows the percentage of nodes present in clusters obtained from the
original and published data. Note that perturbation has little to no
effect over small number of clusters as the distribution of nodes is
identical.
 \begin{figure*}[t]
 \centering
 \subfigure[\scriptsize 2-Clusters]{
 {\includegraphics[width=0.5\columnwidth]{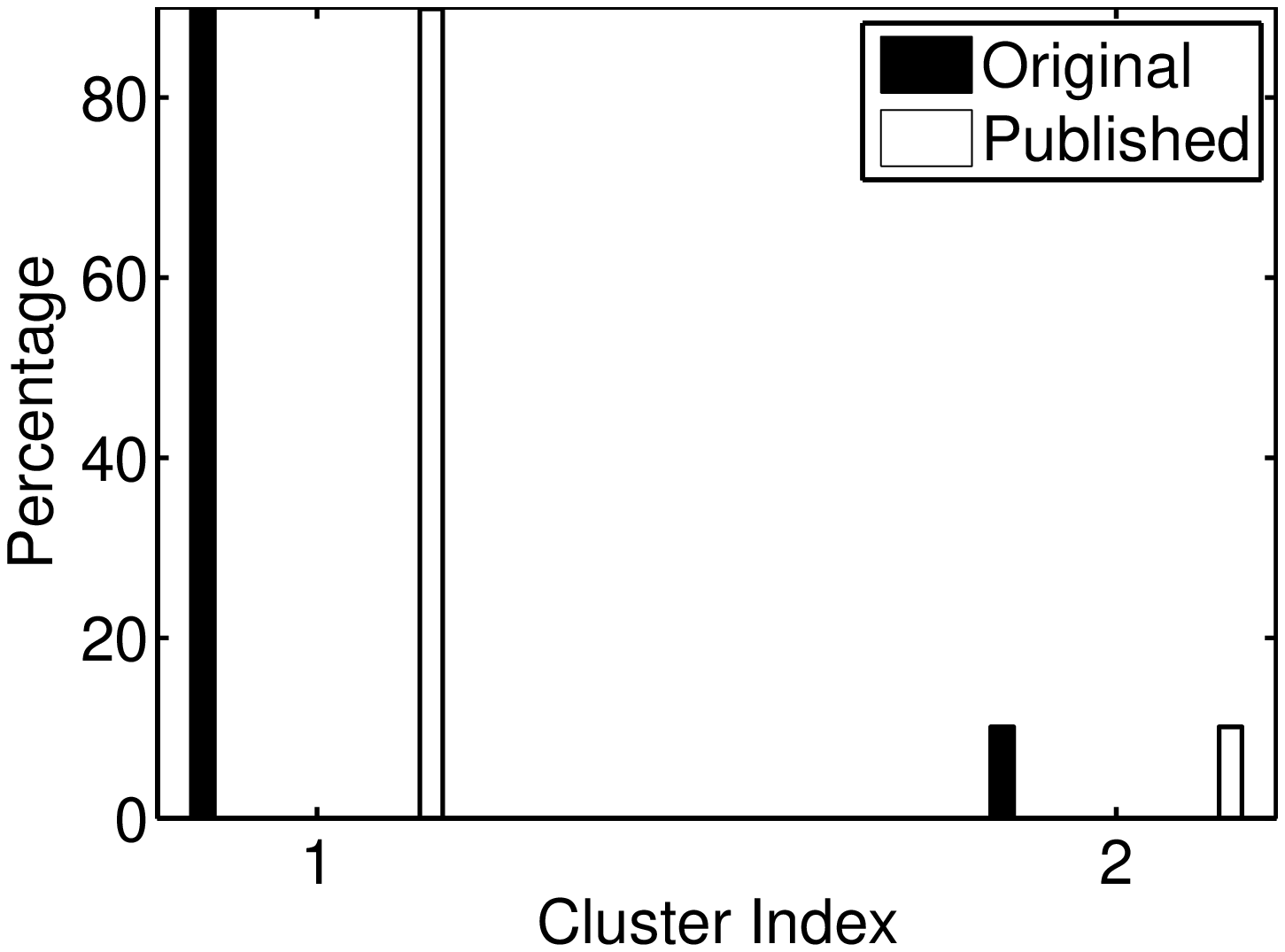}}
 \label{fig:fc2}}
 \hspace{-0.1in}
 \subfigure[\scriptsize 4-Clusters]{
 {\includegraphics[width=0.5\columnwidth]{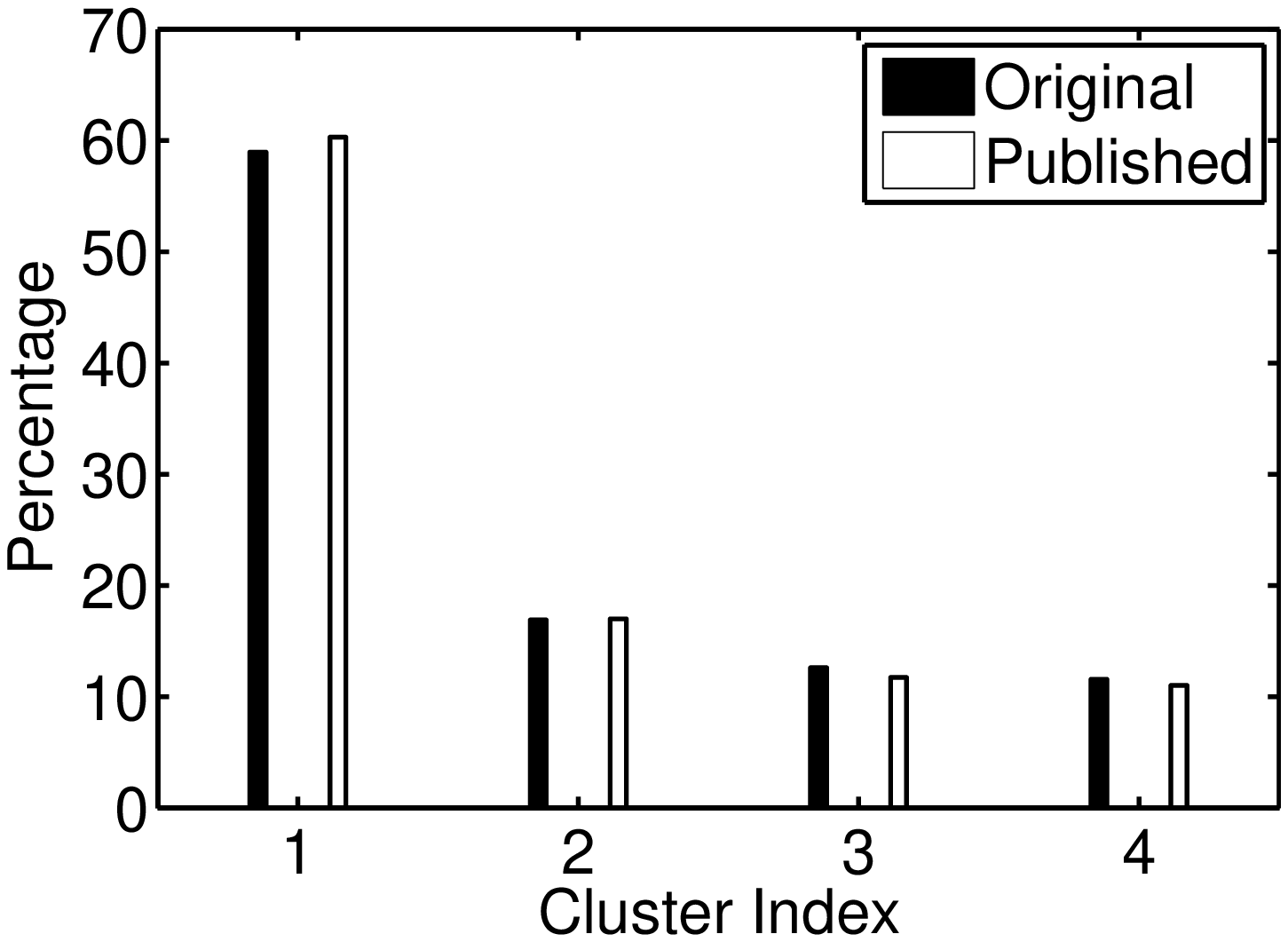}}
 \label{fig:fc4}}
 \hspace{-0.1in}
  \subfigure[\scriptsize 8-Clusters]{
 {\includegraphics[width=0.5\columnwidth]{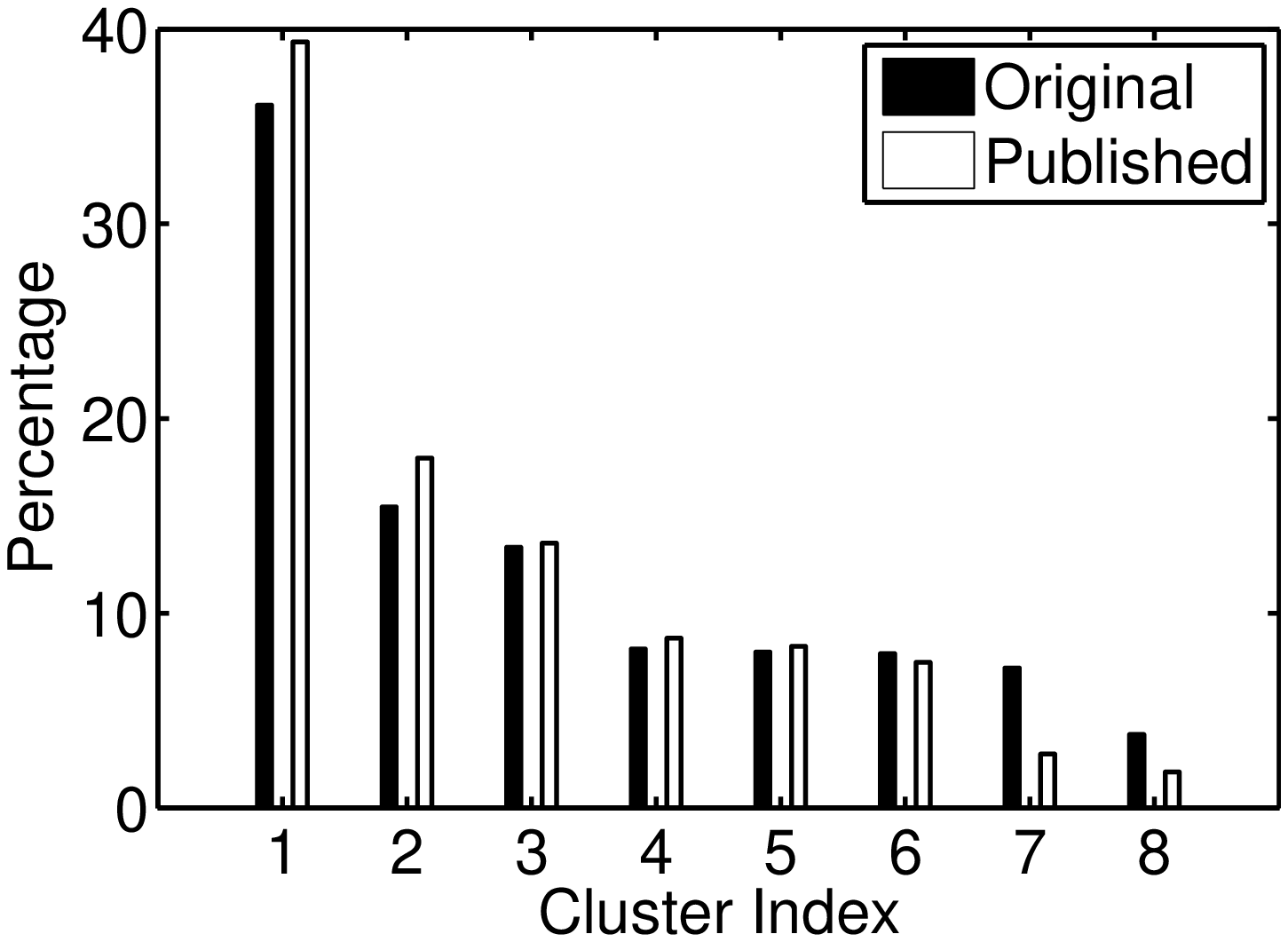}}
 \label{fig:fc8}}
 \hspace{-0.1in}
 \subfigure[\scriptsize 16-Clusters]{
 {\includegraphics[width=0.5\columnwidth]{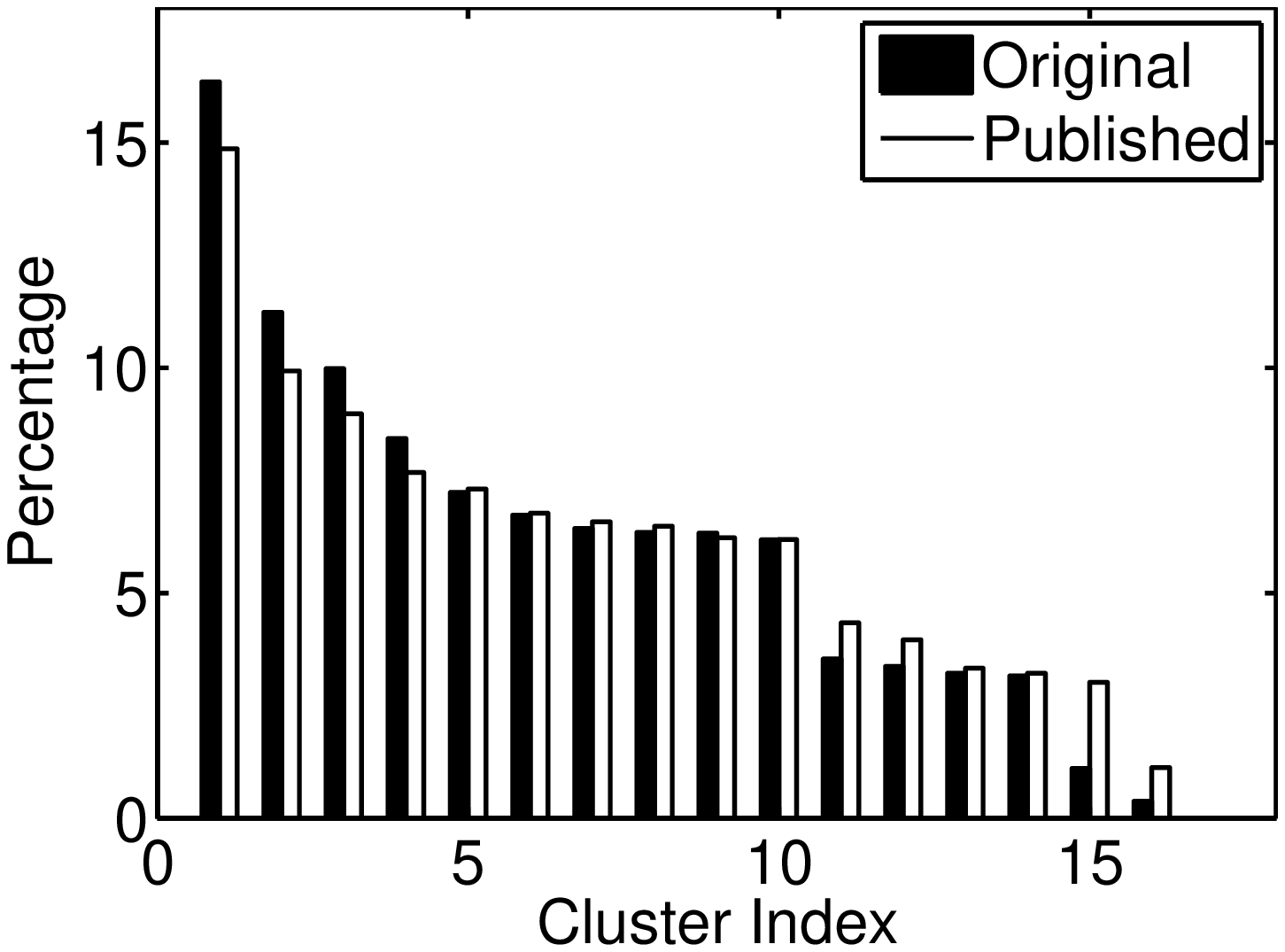}}
 \label{fig:fc16}}
 \caption{Cluster Distribution for Facebook Dataset}
 \label{fig:facebookC}
 \end{figure*}

  \begin{figure*}[!t]
 \centering
 \subfigure[\scriptsize 2-Clusters]{
 {\includegraphics[width=0.5\columnwidth]{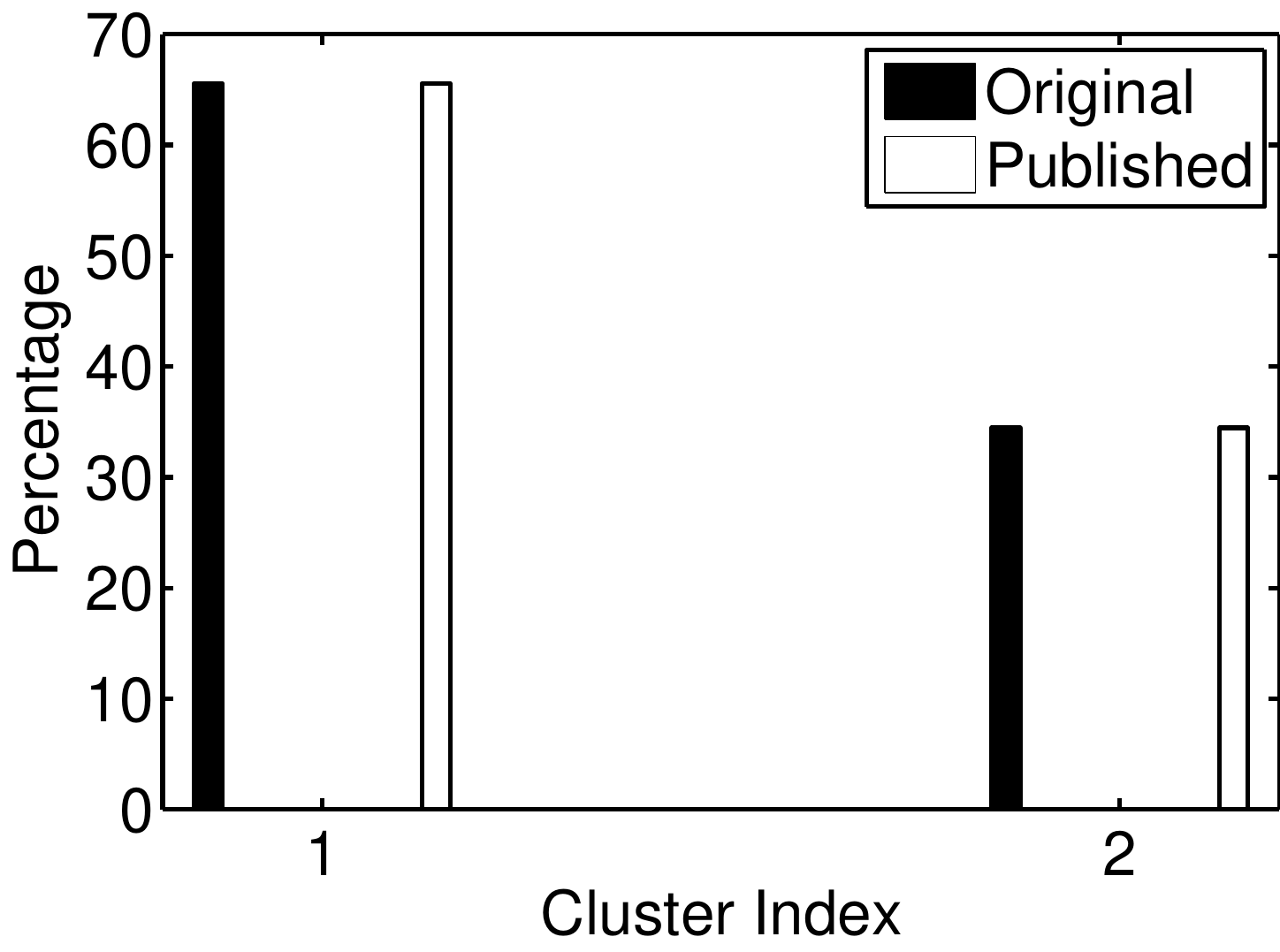}}
 \label{fig:jc2}}
 \hspace{-0.1in}
 \subfigure[\scriptsize 4-Clusters]{
 {\includegraphics[width=0.5\columnwidth]{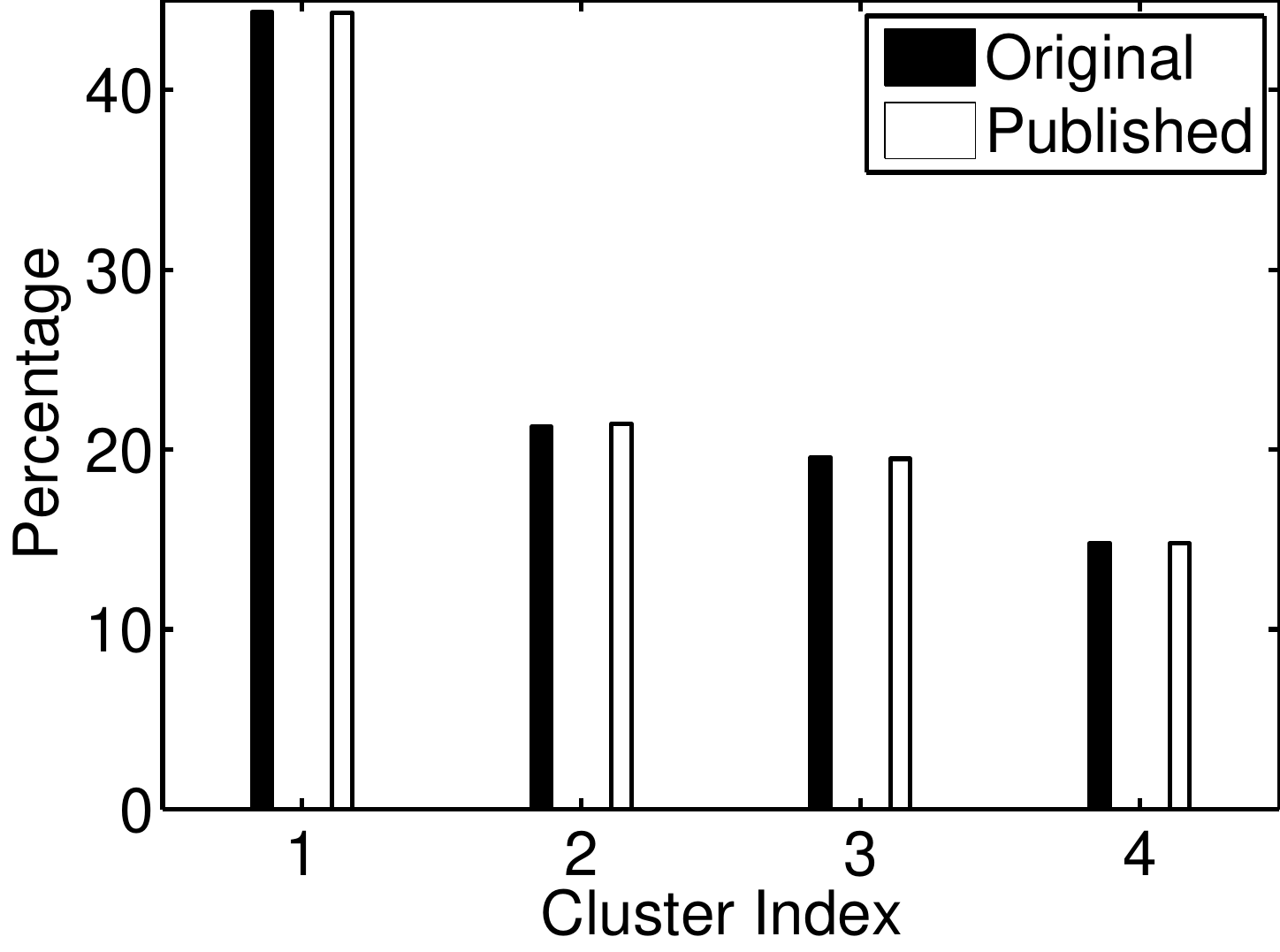}}
 \label{fig:jc4}}
 \hspace{-0.1in}
  \subfigure[\scriptsize 8-Clusters]{
 {\includegraphics[width=0.5\columnwidth]{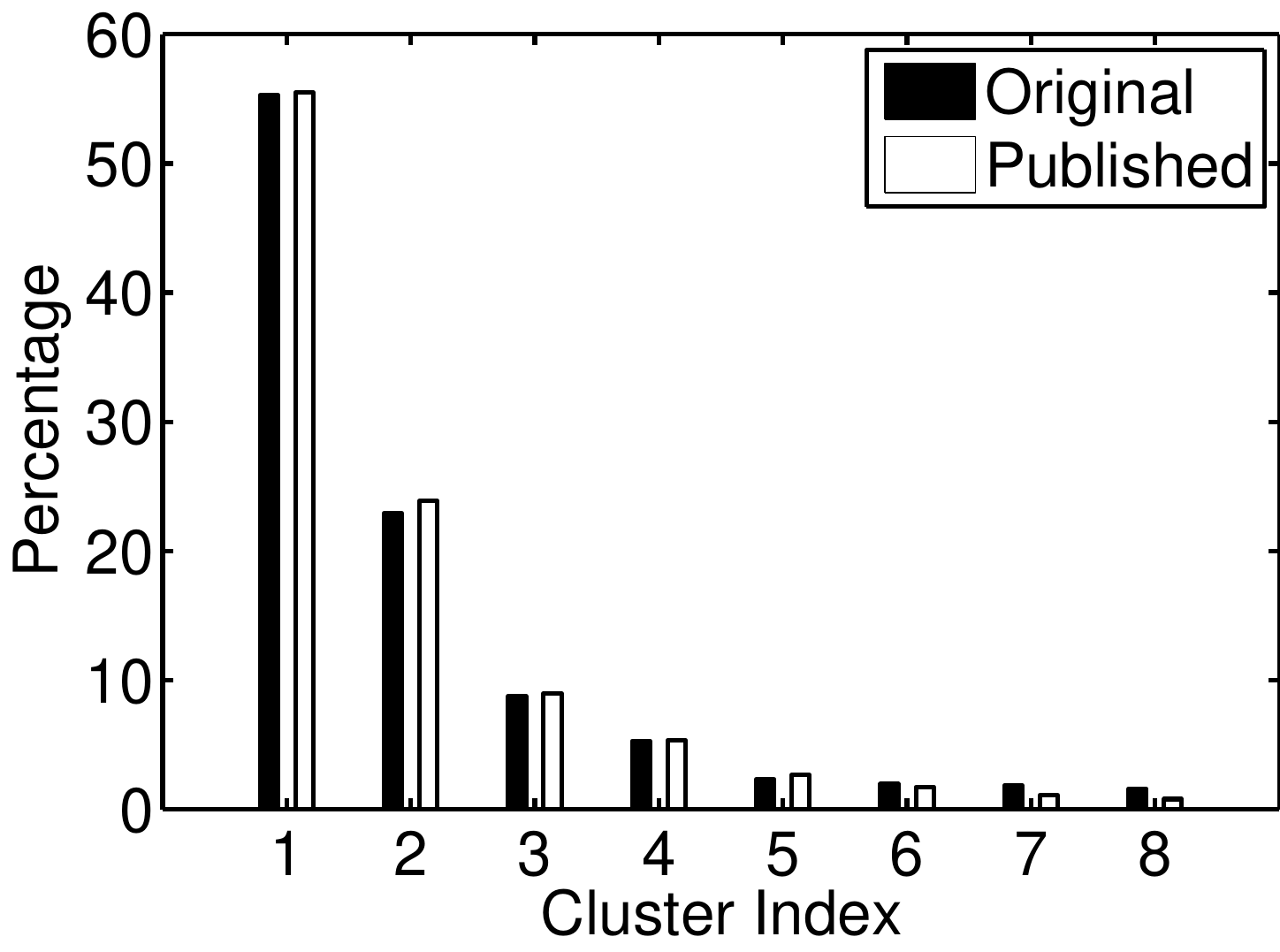}}
 \label{fig:jc8}}
 \hspace{-0.1in}
 \subfigure[\scriptsize 16-Clusters]{
 {\includegraphics[width=0.5\columnwidth]{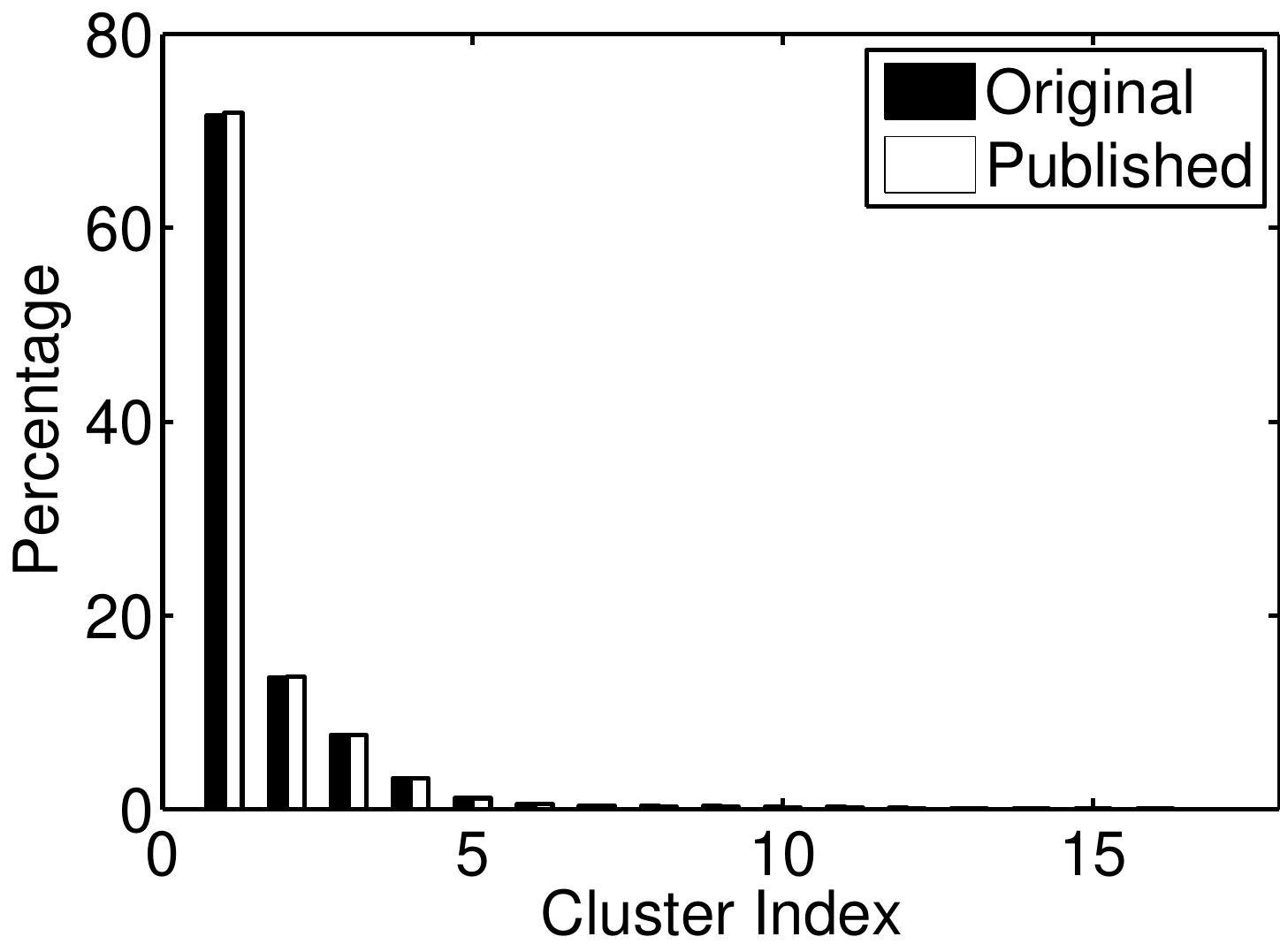}}
 \label{fig:jc16}}
 \caption{Cluster Distribution for Live Journal Dataset}
 \label{fig:journalC}
 \end{figure*}

   \begin{figure*}[!t]
 \centering
 \subfigure[\scriptsize 2-Clusters]{
 {\includegraphics[width=0.5\columnwidth]{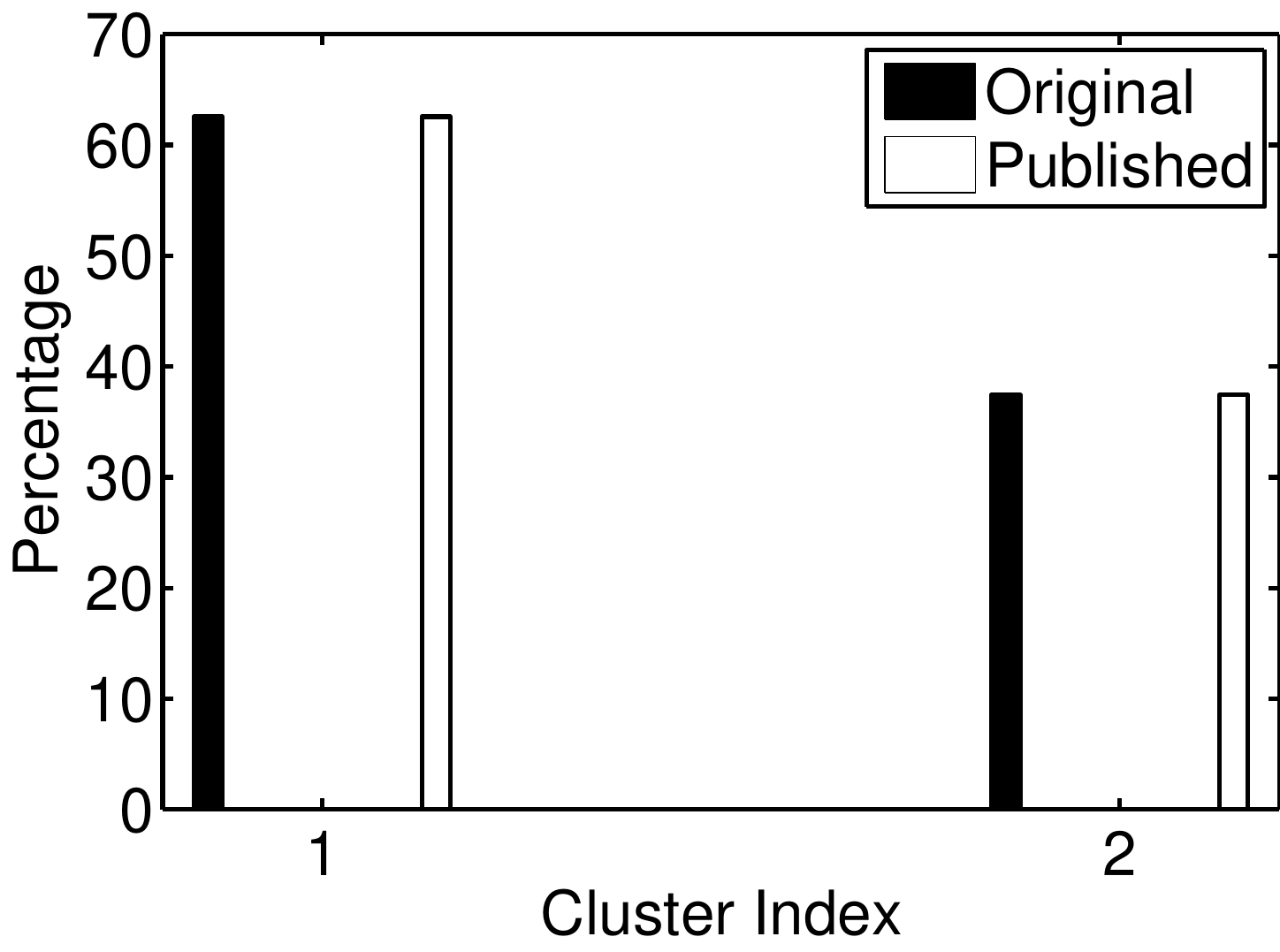}}
 \label{fig:pc2}}
 \hspace{-0.1in}
 \subfigure[\scriptsize 4-Clusters]{
 {\includegraphics[width=0.5\columnwidth]{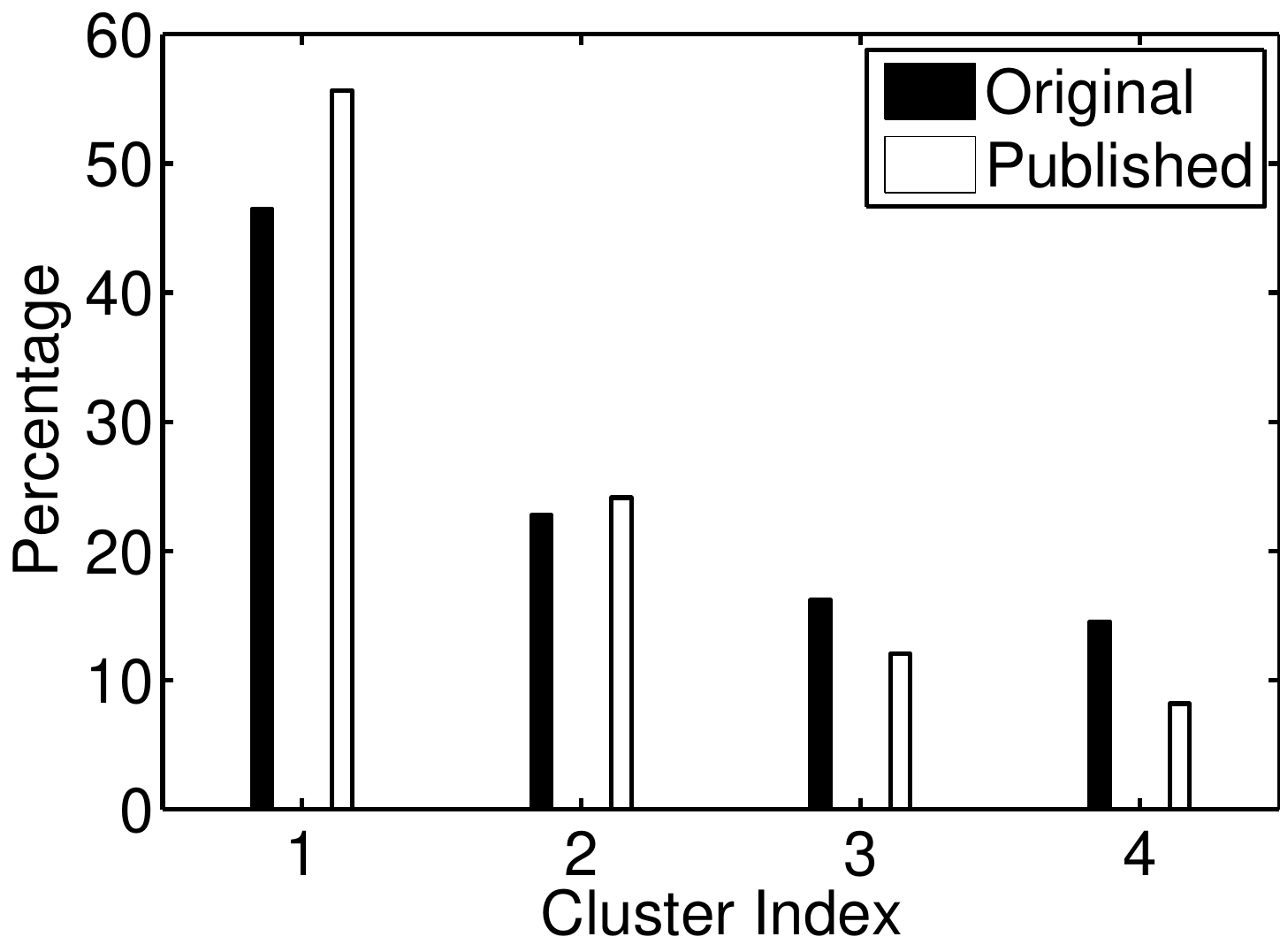}}
 \label{fig:pc4}}
 \hspace{-0.1in}
  \subfigure[\scriptsize 8-Clusters]{
 {\includegraphics[width=0.5\columnwidth]{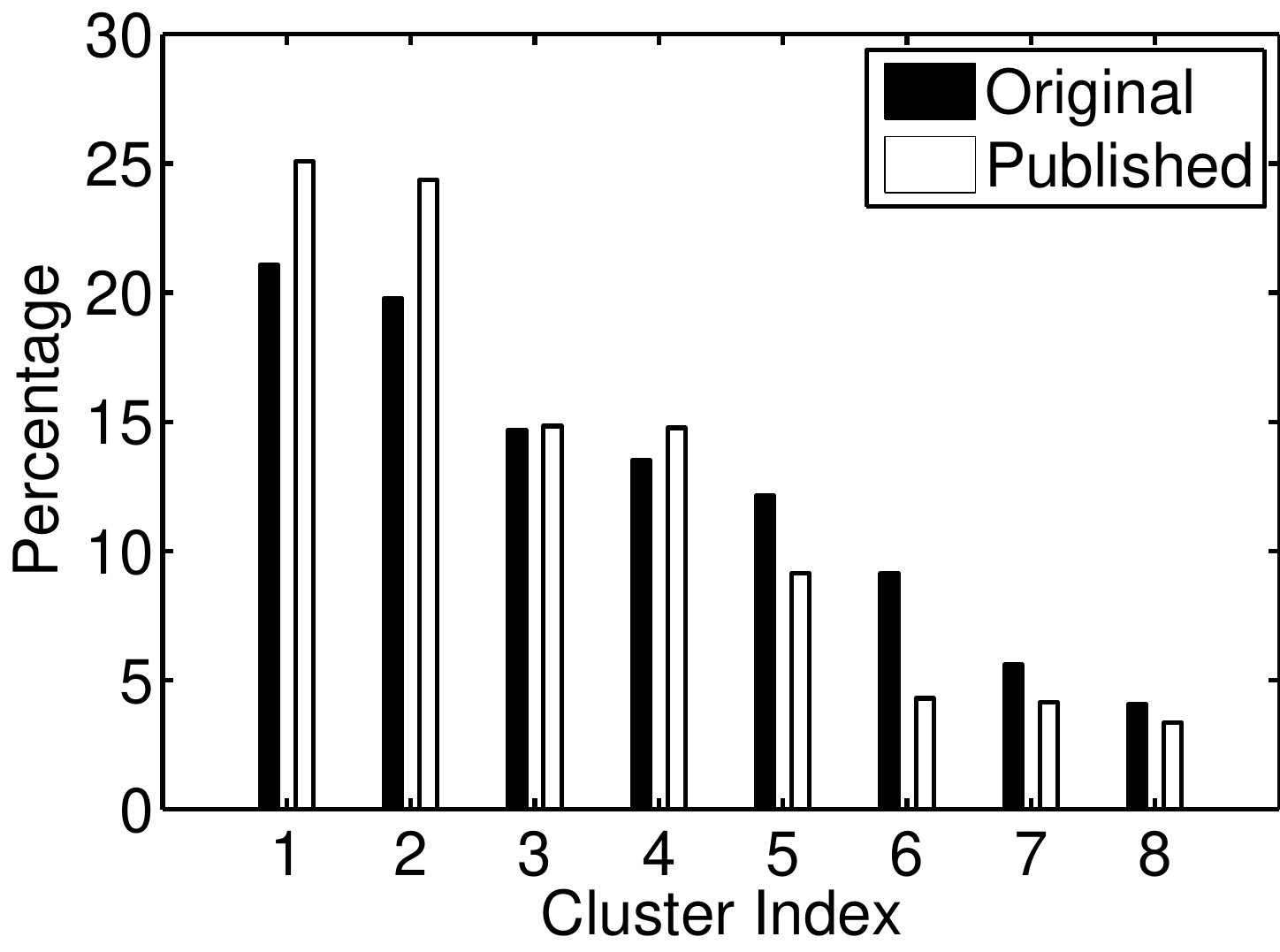}}
 \label{fig:pc8}}
 \hspace{-0.1in}
 \subfigure[\scriptsize 16-Clusters]{
 {\includegraphics[width=0.5\columnwidth]{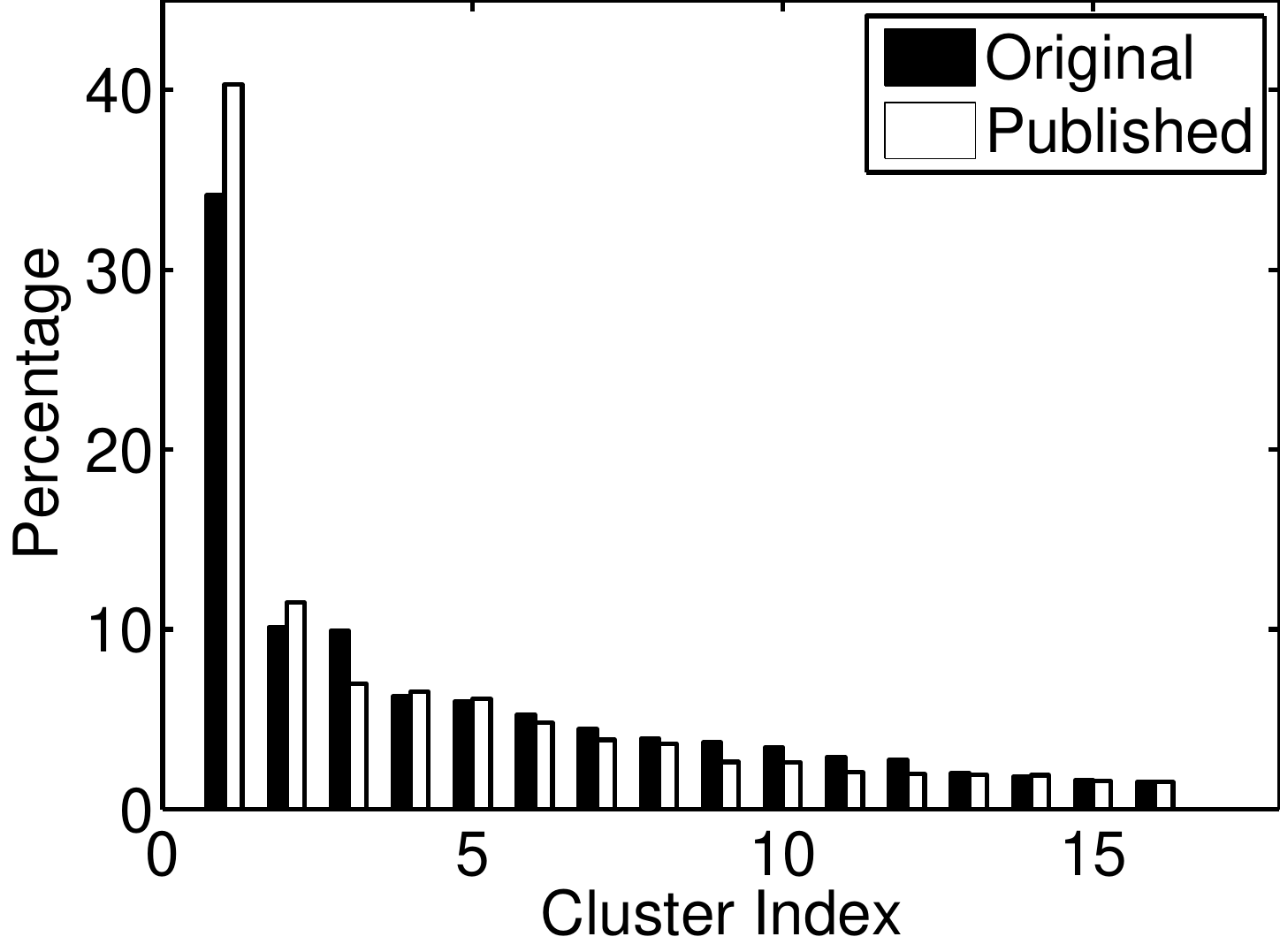}}
 \label{fig:pc16}}
 \caption{Cluster Distribution for Pokec Dataset}
 \label{fig:pokecC}
 \end{figure*}

We compare our results with an approach presented in
\cite{wangdifferential}, which directly perturbs the eigenvector of
the original data by a Laplacian noise. We refer to this approach as
(LNPP) for short. We note that we did not compare to the other
approaches for differential private eigen decomposition because
they are computationally infeasible for the large social networks
studied in our experiments. We implement the LNPP mechanism and
evaluate the clustering performance by comparing it to the
clustering results generated by the original adjacency matrix. Table
\ref{table:LNPPnmi} gives NMI results using LNPP over different
datasets for $\sigma=1$. It is clear that LNPP performs
significantly worse than the proposed algorithm in clustering. Note
that we did not include the clustering performance of LNPP in Figure
\ref{fig:fnmi}, \ref{fig:ljnmi} and \ref{fig:pnmi} because of its poor
performance that basically overlaps with the horizonal axis.

\begin{table}
\centering
\begin{tabular}{|c|c|c|c|}
  \hline
  Dataset & Facebook & Pokec & LiveJournal  \\\hline
  Memory (MB)$m=200$& $4955$ & $2612$ & $6396$  \\\hline
  Memory (MB)$m=20$& $495$ & $261$ & $639$  \\\hline
  Time (sec)$m=200$& $150$ & $97$ & $211$ \\\hline
  Time (sec)$m=20$& $6.15$ & $4.60$ & $8.15$ \\
  \hline
\end{tabular}
\caption{Memory utilization and running time for the proposed algorithm}
\label{table:timesU}
\end{table}

\begin{table}
\centering
\begin{tabular}{|c|c|c|c|c|}
  \hline
  Cluster & $2$ & $4$ & $8$ & $16$ \\\hline
  Facebook& $9.1E-8$ & $8.8E-7$ & $4.1E-6$ & $1.3E-5$  \\\hline
  LiveJournal& $9.7E-7$ & $3.2E-6$ & $3.6E-6$ & $1.1E-5$  \\\hline
  Pokec & $1.1E-7$ & $3.5E-6$ & $5.8E-6$ & $2.6E-5$ \\
  \hline
\end{tabular}
\caption{Clustering result (measured in NMI) using LNPP Approach
\cite{wangdifferential} for $\sigma = 1$ } \label{table:LNPPnmi}
\end{table}

\subsection{Influential Node Analysis} \label{sec:page} Identifying
information hubs in a social network is an important problem.
An information hub refers to a node which occupies a central
position in the community and has a large number of connections with
other users.
Such central nodes play an important role in information diffusion.
Advertising agencies, can utilize information about top-$t$
influential nodes for word-of-mouth advertisements
\cite{ma2008mining}.
Therefore, the preservation of privacy of such influential nodes is
important.

Influential node analysis require information about the
eigen-spectrum of the social network graph.
Eigen-vector centrality (EVC) is a measure to quantify the influence
of a node in a social network \cite{bonacich1972factoring}.
EVC is mathematically related to several other influence measures
such as \cite{katz1953new,taylor1969influence,hoede1978new}.
EVC requires the computation of eigen-vectors and assigns ranks to
nodes according to their location in the most dominant community.
EVC of an adjacency matrix is defined as its principle eigenvector.
We employ principal component centrality (PCC) which is based on EVC
measure to rank the nodes \cite{ilyas2011distributed}.
Let $k$ denote the number of eigen vectors to be computed.
Let $U$ denote the $n\times k$ matrix whose $ith$ column represents
the $ith$ eigenvector of an $n\times n$ adjacency matrix $A$.
Then PCC can be expressed as:
\begin{equation}\label{eq:Pcc}
    C_k = \sqrt{((AU_{n\times k})\bigodot(AU_{n\times k})1_{k\times 1}}
\end{equation}
Where $C_k$ is an $n\times 1$ vector containing PCC score of each
node.
Nodes with highest PCC scores are considered the influential
nodes.
Similar to the clustering approach, Algorithm \ref{alg:PCC} gives
the standard PCC algorithm, and Algorithm \ref{alg:diffPcc} states
the key steps of differential private PCC algorithm.
\begin{algorithm}
\small \LinesNumbered \DontPrintSemicolon \KwIn{(1) Adjacency
Matrix $A\in \mathbb R^{n\times n}$\\
\hspace{0.41in}(2) number of top eigenvectors $k$\\
} \KwOut{PCC score $C_k$}

\BlankLine Compute first $k$ eigenvectors $\u_1,..,\u_k$ of $A$\;
Get matrix $U\in \mathbb R^{n\times k}$ where $ith$ column of $U$ is
$\u_i$\; Obtain PCC scores $C_k$ using Equation \ref{eq:Pcc}\;
\caption{\textbf{Principal Component Centrality}}\label{alg:PCC}
\end{algorithm}

\begin{algorithm}
\small \LinesNumbered \DontPrintSemicolon \KwIn{(1) adjacency
matrix $A\in \mathbb R^{n\times n}$\\
\hspace{0.41in}(2) number of top eigenvectors $k$\\
\hspace{0.41in}(3) the number of random projections $m < n$\\
\hspace{0.41in}(4) variance for random noise $\sigma^2$\\
} \KwOut{PCC score $\hat{C}_k$}

\BlankLine {Compute a differential private matrix for social network
$A$ by $\widehat{A} = \texttt{Publish}(A, m, \sigma^2)$} \; Compute
first $k$ eigenvectors $\ut_1,..,\ut_k$ of $\widehat{A}$\; Get
matrix $U\in \mathbb R^{n\times k}$ where $ith$ column of $U$ is
$\ut_i$\; Obtain PCC scores $\hat{C}_k$ using Equation
\ref{eq:Pcc}\; \caption{\textbf{Differential Private Principal
Component Centrality}}\label{alg:diffPcc}
\end{algorithm}

We evaluate the utility preservation of the published data by
evaluating the accuracy with which influential nodes with high ranks
are identified. First, for a given value of $k$, eigenvectors
corresponding to the $k$ largest eigenvalues are computed from the
original adjacency matrix and used to obtain PCC scores of all the
nodes in a graph using Algorithm \ref{alg:PCC} (denoted as $C_k$).
Then, a second set of $k$ eigenvectors is computed from the
published data i.e., after applying matrix randomization using
Algorithm \ref{alg:diffPcc}. This second set is then used to obtain
another vector containing PCC scores denoted as $\hat{C}_k$. The
original scores $C_k$ and the published scores $\hat{C}_k$ are then
compared in two different ways. For all experiments, we compute PCC
scores by varying the number of eigenvectors in the range
$k=2,4,8,16$.

In the first evaluation, we use Mean Square Error (MSE) to compute
the error between score values of $C_k$ and $\hat{C}_k$. We report
$n\times MSE$ in our study in order to alleviate the scaling factor
induced by the size of social networks. In the second evaluation, we
identify two sets of top $t$ influential nodes based on the PCC
scores computed from the original data as well as from the published
data. We then evaluate the performance of our algorithm by measuring
the percentage of overlapped nodes between these two sets. Table
\ref{table:ranks} gives the values of Mean Square Error between PCC
scores obtained from the original and published data. We also
compare these results with the LNPP approach. For comparison, we
show in Table \ref{table:pccL} the MSE results for baseline LNPP. It
is clear that the proposed algorithm yields significantly more
accurate estimation of PCC scores than LNPP. In most cases, the
proposed approach is $100$ times more accurate than LNPP.


\begin{table}[!t]
\centering
\begin{tabular}{|c|c|c|c|c|}
  \hline
  \# of Eigenvectors & $2$ & $4$ & $8$ & $16$ \\\hline
    Facebook & $2.6e^{-26}$ & $2.9e^{-4}$ & $0.021$ & $0.013$\\ \hline
    Live Journal & $4.0e^{-4}$ & $0.006$ & $0.034$ &$0.719$\\ \hline
    Pokec & $3.0e^{-4}$ & $0.005$ & $0.009$ &$0.019$\\
  \hline
\end{tabular}
\caption{$n\times \mbox{MSE}$ using the proposed approach} \label{table:ranks}
\end{table}

\begin{table}[!t]
\centering
\begin{tabular}{|c|c|c|c|c|}
  \hline
  \# of Eigenvectors & $2$ & $4$ & $8$ & $16$ \\\hline
    Facebook & $1.83$ & $1.83$ & $1.67$ & $1.64$\\ \hline
    Live Journal & $1.96$ & $1.96$ & $1.88$ &$1.92$\\ \hline
    Pokec & $1.79$ & $1.63$ & $1.62$ &$1.55$\\
  \hline
\end{tabular}
\caption{$n\times \mbox{MSE}$ using baseline LNPP} \label{table:pccL}
\end{table}

In the second evaluation, we measure the percentage of nodes correctly identified as the
top$-t$ influential nodes. First, we obtain two sets $T$ and $\hat{T}$ that contain the top$-t$ most influential nodes measured by the PCC scores given by $C_k$ and $\hat{C}_k$. Then the percentage of nodes common to both $T$ and $\hat{T}$ is computed. We consider top $10,100,1000$ and $10000$ ranked nodes.
Figure \ref{fig:percMR} shows the percentage of nodes correctly identified as the top$-t$ influential nodes for the three datasets. Figure \ref{fig:percL} gives the results for LNPP approach. We can see that for all case, the proposed algorithm is able to recover at least $80\%$ of the most influential nodes.
In contrast, LNPP fails to preserve the most influential nodes as the percentage
of nodes correctly identified as the top$-t$ influential nodes is
less than $1\%$ for all cases.

 \begin{figure*}[t]
 \centering
 \subfigure[\scriptsize Facebook]{
 {\includegraphics[width=0.6\columnwidth]{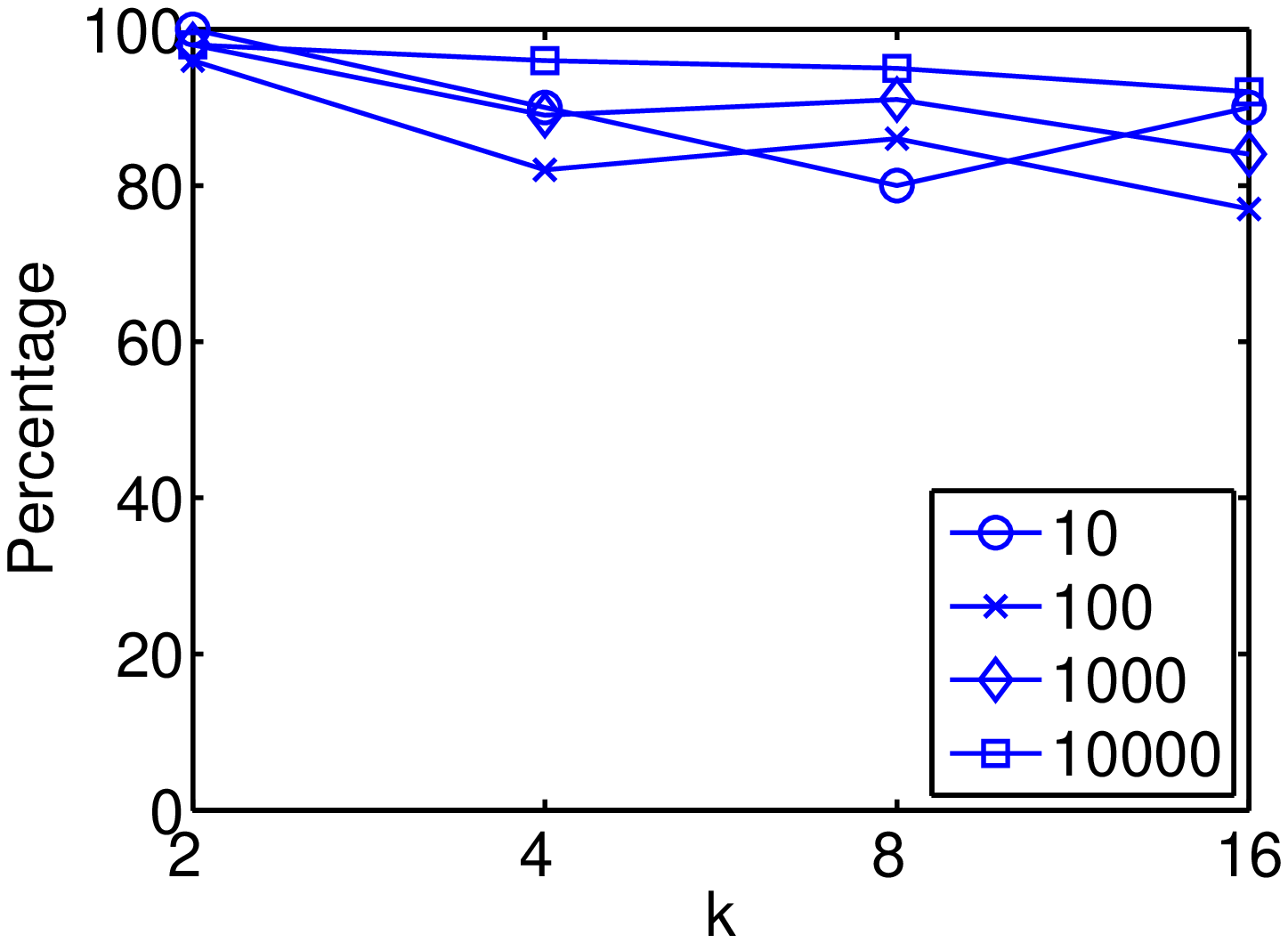}}
 \label{fig:fbm}}
 \hspace{-0.1in}
 \subfigure[\scriptsize LiveJournal]{
 {\includegraphics[width=0.6\columnwidth]{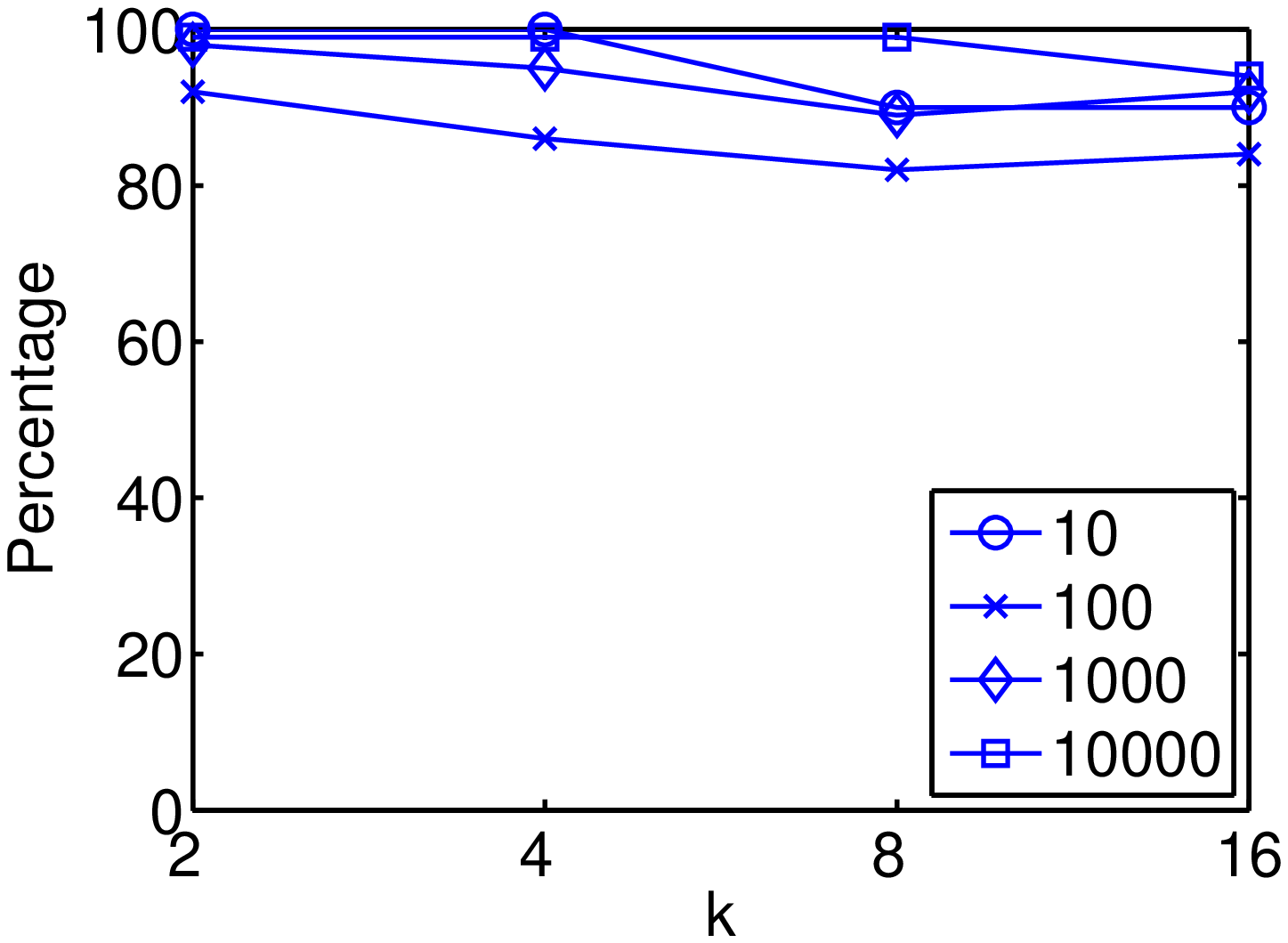}}
 \label{fig:ljm}}
 \hspace{-0.1in}
  \subfigure[\scriptsize Pokec]{
 {\includegraphics[width=0.6\columnwidth]{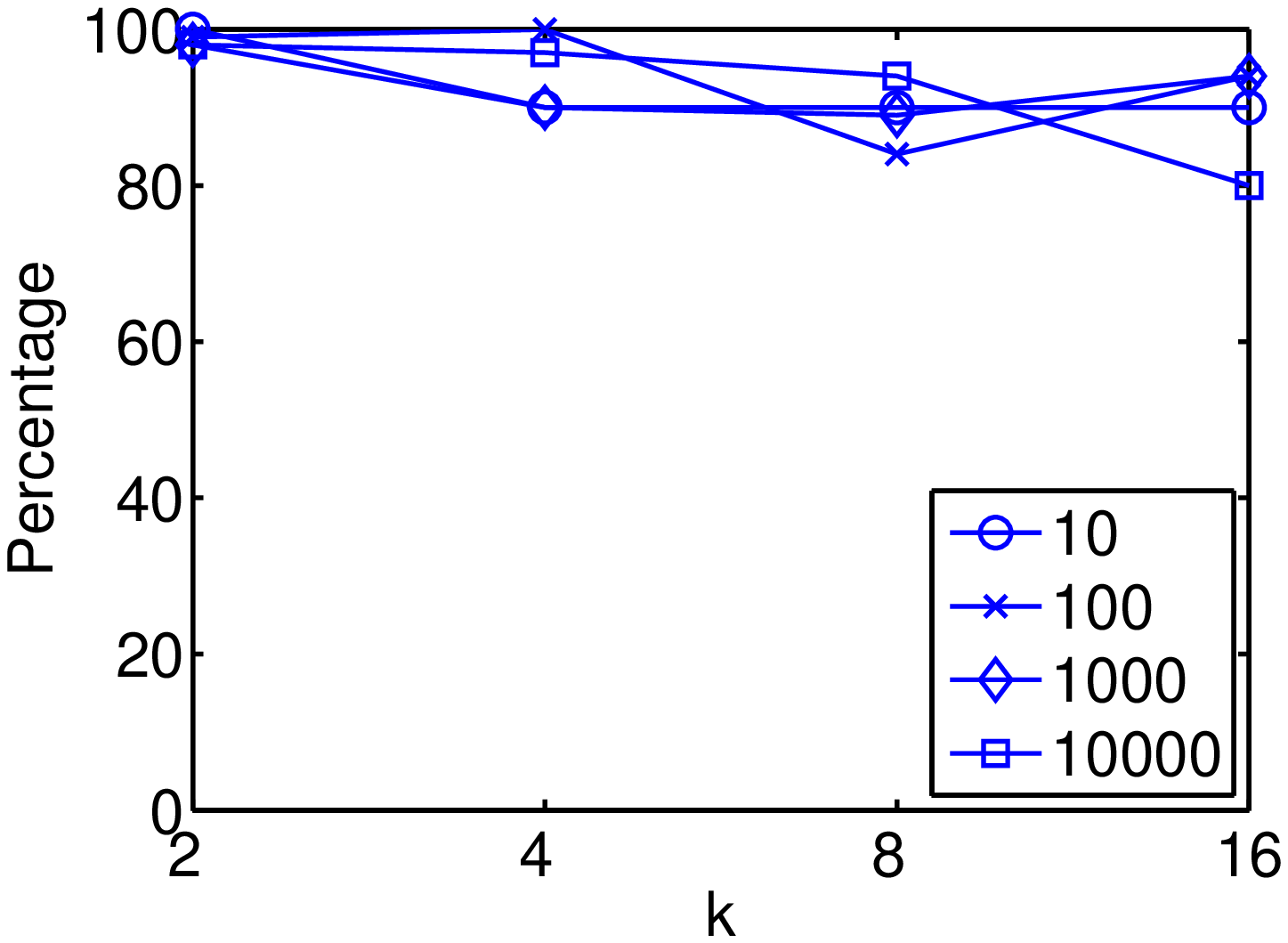}}
 \label{fig:pkm}}
 \caption{Percentage of preserved ranks using random matrix approach}
 \label{fig:percMR}
 \end{figure*}

  \begin{figure*}[t]
 \centering
 \subfigure[\scriptsize Facebook]{
 {\includegraphics[width=0.6\columnwidth]{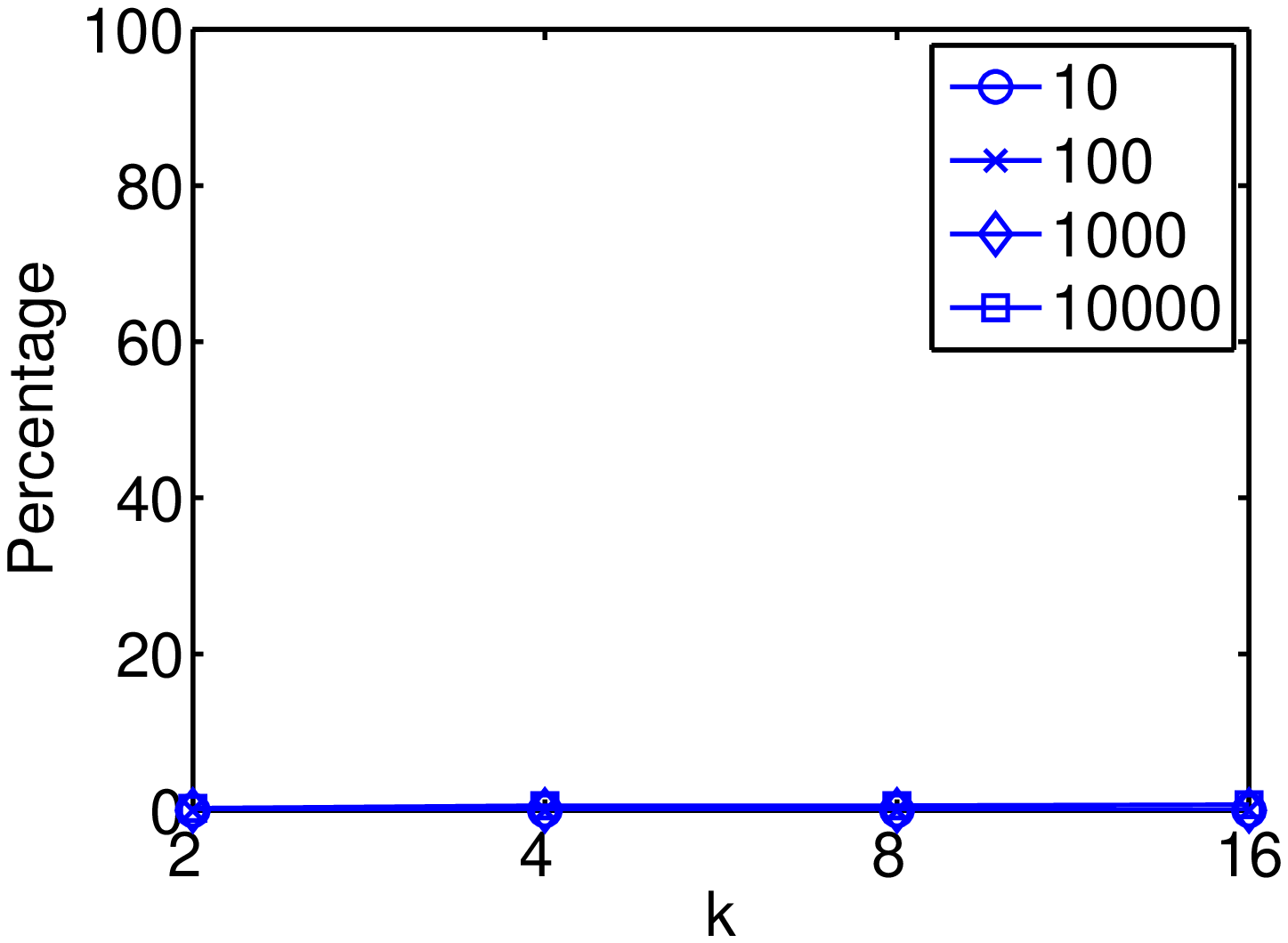}}
 \label{fig:fbl}}
 \hspace{-0.1in}
 \subfigure[\scriptsize LiveJournal]{
 {\includegraphics[width=0.6\columnwidth]{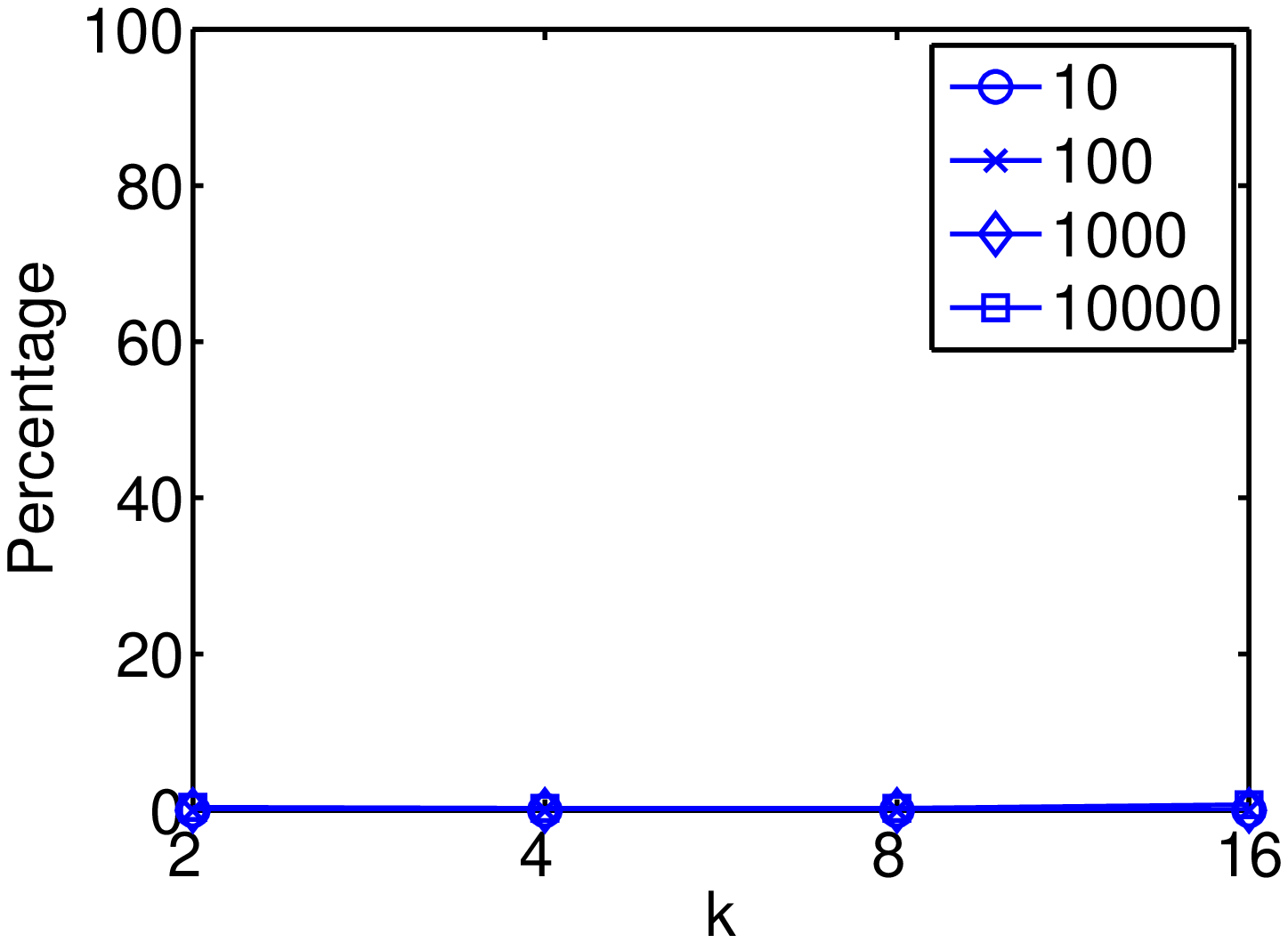}}
 \label{fig:ljl}}
 \hspace{-0.1in}
  \subfigure[\scriptsize Pokec]{
 {\includegraphics[width=0.6\columnwidth]{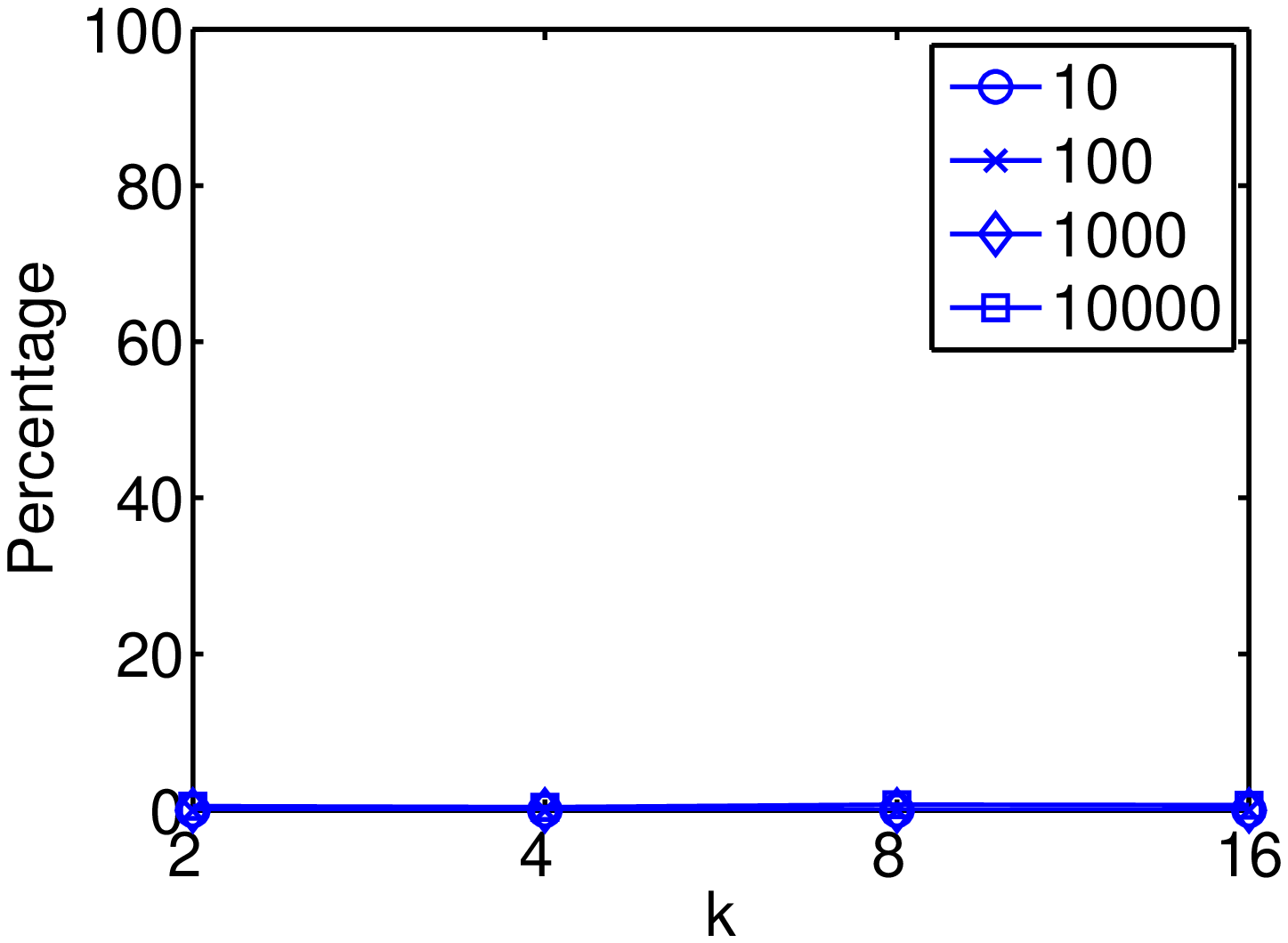}}
 \label{fig:pkl}}
 \caption{Percentage of preserved ranks using LNPP approach}
 \label{fig:percL}
 \end{figure*}

\section{Conclusion}
\label{sec:conclusion} Graphs obtained from large social networking
platforms can provide valuable information to the research
community.
Public availability of social network graph data is problematic due
to the presence of sensitive information about individuals present
in the data.
In this paper present a privacy preserving mechanism for publishing
social network graph data which satisfies differential privacy
guarantees.
We present a random matrix approach which can be utilized for
preserving the eigen-spectrum of a graph.

The random projection approach projects the adjacency matrix $A$ of
a social network graph to lower dimensions by multiplying $A$ with a
random projection matrix $P$.
This approach satisfies differential privacy guarantees by
randomization and maintains utility by adding low level of noise.
For evaluation purposes we use three different social network graphs
from Fcaebook, Live Journal and Pokec.
We analyze the impact of our perturbation approach by evaluating the
utility of the published data for two different applications which
require spectral information of a graph.

We consider clustering of social networks and identification of
influential nodes in a social graph.
The results show that even for high values of noise variance
$\sigma=1$ the clustering quality given by NMI values is as low as
$0.74$
For influential node discovery, the propose approach is able to correctly recover at $80\%$ of the most influential nodes.

}


\bibliographystyle{abbrv}
\bibliography{sig-alternate}
\end{document}